\documentclass[mathpazo]{cicp}

\usepackage{chemformula} 
\usepackage[T1]{fontenc} 
\usepackage{amsmath,amsthm,amssymb}
\usepackage{mathrsfs}
\usepackage{subfigure}
\usepackage{booktabs}
\usepackage{multirow} 
\usepackage{multicol}
\usepackage{float}
\usepackage{bm}

\usepackage[ruled,linesnumbered]{algorithm2e}
\usepackage{ragged2e}

\usepackage{listings}
\usepackage{xcolor}
\begin{document}

\title[Random batch molecular dynamics]{RBMD: A molecular dynamics package enabling to simulate 10 million all-atom particles in a single graphics processing unit}

%
%

\author[Gao W H et~al.]{Weihang Gao\affil{1,3},
      Teng Zhao\affil{2,3}, Yongfa Guo\affil{3}, Jiuyang Liang\affil{1}, Huan Liu\affil{3}, Maoying Luo\affil{3}, Zedong Luo\affil{3}, Wei Qin\affil{3},  Yichao Wang\affil{4}, Qi Zhou\affil{1}, Shi Jin\affil{1,2,3}, and Zhenli Xu\affil{1,2,5}\comma\corrauth}
\address{\affilnum{1}\ School of Mathematical Sciences, Shanghai Jiao Tong University, Shanghai 200240, China \\
          \affilnum{2}\ Institute of Natural Sciences, MOE-LSC, and Shanghai National Center for Applied Mathematics, Shanghai Jiao Tong University, Shanghai 200240, China \\
          \affilnum{3}\ Shanghai Jiao Tong University-Chongqing Institute of Artificial Intelligence, Chongqing 401329, China\\ 
          \affilnum{4}\ Network $\&$ Information Center, Shanghai Jiao Tong University, Shanghai 200240, China\\
          \affilnum{5}\ CMA-Shanghai, Shanghai Jiao Tong University, Shanghai 200240, China}
\email{{\tt xuzl@sjtu.edu.cn} (Zhenli Xu)}

\begin{abstract}
This paper introduces a random-batch molecular dynamics (RBMD) package for fast simulations of particle systems at the nano/micro scale. Different from existing packages, the RBMD uses random batch methods for nonbonded interactions of particle systems. The long-range part of Coulomb interactions is calculated in Fourier space by the random batch Ewald algorithm, which achieves linear complexity and superscalability, surpassing classical lattice-based Ewald methods. For the short-range part, the random batch list algorithm is used to construct neighbor lists, significantly reducing both computational and memory costs.  
The RBMD is implemented on GPU-CPU heterogeneous architectures, with classical force fields for all-atom systems. Benchmark systems are used to validate accuracy and performance of the package.  Comparison with the particle-particle particle-mesh method and the Verlet list method in the LAMMPS package is performed on three different NVIDIA GPUs, demonstrating high efficiency of the RBMD on heterogeneous architectures.  Our results also show that the RBMD enables simulations on a single GPU with a CPU core up to 10 million particles. Typically, for systems of one million particles, the RBMD allows simulating all-atom systems with a high efficiency of $8.20$ ms per step, demonstrating the attractive feature for running large-scale simulations of practical applications on a desktop machine. 

	
\end{abstract}
\keywords{Molecular dynamics; random batch methods; heterogeneous architectures; Coulomb interactions; large-scale simulations}
\maketitle
%
%
%


{\bf PROGRAM SUMMARY}

\begin{small}
\noindent
{\em Program Title: }RBMD                                  \\
{\em Developer's repository link:} https://github.com/randbatch-md\\
{\em Licensing provisions:} GPL 3.0 \\
{\em Programming language: }C++                               \\
{\em Nature of problem: }Nonbonded interactions, particularly,  long-range electrostatic interactions, are the computational bottleneck of molecular dynamics simulations, for which most packages are based on algorithms with the lattice-based Ewald summation and the fast Fourier transform. These algorithms are  communication intensive, limiting the parallel efficiency of the algorithm for the GPU calculation. Moreover, for short-range calculations, the building of neighbor lists for each particle is memory intensive, leading to the von Neumann bottleneck for large scale simulations. Therefore, the use of innovative long-range and short-range algorithms is essential for a novel molecular dynamics package that is able to simulate systems beyond the current limitations. \\
{\em Solution method:}  In the paper, the random batch methods are introduced under the VTK-m framework to accelerate the nonbonded interactions in molecular dynamics, leading to a novel package, named the random batch molecular dynamics (RBMD). The stochastic nature of the methods significantly improves the efficiency and parallel scalability of the calculations on the GPU-CPU heterogeneous architecture. \\
\\

\end{small}


\section{Introduction}

Over the past few decades, molecular dynamics (MD) simulations have achieved tremendous success in a broad range of areas, including biophysics, soft matter, materials modeling, electrochemical energy devices and drug design \cite{karplus1990molecular,Axel1987Science,morgane2018polymer,CiCP-33-57}. These advances  partly owe to tremendous improvements in hardware~\cite{leiserson2020there} that have enabled studies of spatial and temporal scales previously not feasible, such that a more detailed understanding of physical phenomena at micro/macro-scales can be achieved. At present, most traditional MD packages have released  multi-CPU and/or GPU versions, such as AMBER~\cite{amber18}, GROMACS~\cite{gromacsJCP2020}, LAMMPS~\cite{THOMPSON2022108171}, NAMD~\cite{phillips2020scalable}, and OpenMM~\cite{OpenMM2017}. Moreover, the development of accelerators~\cite{jones2022accelerators} has also promoted the development of machine learning-integrated MD packages \cite{huang2022sponge,wang2018deepmd,fan2022gpumd}.
Nevertheless, since Moore's law~\cite{leiserson2020there} is no longer applicable as the speed improvement of individual processors has been increasing slowly, fast and accurate calculations of pairwise nonbonded interactions on heterogeneous architectures have become a fundamental challenge in the field. This is because atomic Coulomb interactions~\cite{RevModPhys.82.1887} are long-range and require costly all-to-all communications with the increase of CPU cores or GPUs. Mainstream MD software packages typically address Coulomb interactions through Ewald-type methods~\cite{Hockney1988Computer,darden1993particle,essmann1995smooth,deserno1998mesh,dos2016simulations} where the long-range part is accelerated using the fast Fourier transform (FFT). These methods achieve a complexity of $O(N\log N)$, but the intensive communications required by the FFT significantly reduce the parallel scalability~\cite{ayala2021scalability}. Other popular Coulomb solvers~\cite{greengard1987fast,barnes1986hierarchical,maggs2002local,pei2023fast} including the fast multipole method also suffer from the same  issue~\cite{arnold2013comparison}.

Another critical challenge limiting the spatio-temporal scale of modern MD software is the von Neumann bottleneck~\cite{backus1978can}. This issue arises from the disparity between the processing speed of the CPU/GPU and the data transfer rate between the processor and memory. In classical MD simulations, calculating short-range interactions is particularly data-intensive, requiring large memory consumption and frequent accesses to store and retrieve particles and their neighbors. Without algorithms that can better exploit new hardware architectures and parallel frameworks, the accessible timescales in simulations will quickly hit a wall with the increase in system size.

Stochastic algorithms have emerged gradually, building an important bridge linking traditional methods and modern massive high-performance computing. Notably, the random batch Ewald (RBE)~\cite{jin2021random} and random batch list (RBL)~\cite{liang2021random} methods recently proposed by some of us have shown promise in addressing the aforementioned challenges associated with long-range and short-range interactions, respectively~\cite{liang2022superscalability,liang2022random}. The so-called ``random mini-batch'' idea adopted in the RBE and RBL has its origin in the stochastic gradient descent~\cite{robbins1951textordfemininea}, first proposed for interacting particle systems with rigorous error estimates~\cite{jin2020random,jin2021convergence,jin2022random}, and has succeeded in various fields~\cite{li2020random,carrillo2021consensus,JinJCM-39-897,CiCP-28-1907,CiCP-31-997}. 
The RBE method is based on the Ewald summation~\cite{Ewald1921AnnPhys}, but avoids the use of FFT by employing a random batch importance sampling strategy in the Fourier space calculations, achieving both a near-optimal $O(N)$ complexity and a substantial reduction in communication cost. The RBL method replaces the traditional Verlet list~\cite{PhysRev.159.98} by dividing the region into a core-shell structure and constructing the mini-batch for particles in the shell region, significantly reducing the memory usage and computational cost. Although CPU parallelization of these stochastic algorithms has been widely tested~\cite{liang2022superscalability}, no GPU-involved implementation has been reported to date, and they have yet to be incorporated into any publicly released MD package.

In this paper, we present a random batch molecular dynamics (RBMD) package designed for efficient and scalable MD simulations. The RBMD integrates a CPU-GPU heterogeneous implementation of both the RBE and RBL methods for accelerating nonbonded interactions. It supports mainstream force fields, thermostats, and constraint algorithms, enabling it to accommodate various user requirements. Although heterogeneous offload techniques have been explored by several MD codes~\cite{THOMPSON2022108171,gromacsJCP2020}, the RBMD incorporates stochastic algorithms adapted to modern architectures, so the load balancing strategy requires being specially designed. The RBMD is developed under the VTK-m  framework~\cite{bolstad2023vtk}, designing it into a future-proof form to exploit both CPU and GPU (also including various accelerator architectures) to achieve high performance. In the RBMD, most MD procedures are offloaded to accelerators, while only tasks requiring fine-grained parallelism remain on the CPU. One of the main features of the RBMD is that it does not use grid-based computations. Instead, it designs the random mini-batch steps and nonbonded interaction calculations to be heterogeneous and finally employs a set of \textit{Reduce-by-Key} operators to reduce these random forces. By avoiding grid generation and complete neighbor lists, our RBMD software significantly reduces memory consumption while maintaining high computational efficiency. The RBMD enables simulations of all-atom systems with up to $10^7$ particles on a single GPU and one CPU core workstation, achieving performance of $1-2$ orders of magnitude faster than the famous LAMMPS software; thus, it will be promising for practical simulations of many systems at the nano/micro scale.

The paper is organized as follows: Firstly, we introduce the workflow of the RBMD and review the derivation of the RBE and RBL algorithms for nonbonded forces in Section \ref{methods}. Next, we present the framework for GPU programming in the RBMD and describe the parallel operator and implementation of the RBE/RBL algorithms on GPU in Section \ref{gpu_imple}. Finally, we verify the accuracy and efficiency of the RBMD by presenting numerical examples, demonstrating the advantages of the RBMD in Section \ref{discuss}. Concluding remarks are provided in Section \ref{sec::Conclusion}.

\section{Methods}\label{methods}

In this section, we describe the workflow of the RBMD package and the state-of-the-art algorithms used in the package, particularly the nonbonded forces of interacting particles.

\subsection{RBMD workflow}\label{workflow}

MD simulation is performed by solving the Newton's equations for all particles in a system to obtain the trajectories of particle velocities and positions, leading to the thermodynamics and dynamical quantities by ensemble average. The crucial part of the MD simulation is the calculation of the interacting forces of particles. Consequently, the force field, which defines the form of the potential functions and their parameters, becomes the core of MD simulations. Many force fields have been developed and popularly used in practical simulations, including  AMBER \cite{wang2004development}, CHARMM\cite{brooks1983charmm}, GROMOS\cite{oostenbrink2004biomolecular}, OPLS\cite{jorgensen1996development}, COMPASS\cite{sun1998compass}, and ReaxFF\cite{senftle2016reaxff}, together with many recent progresses in machine-learning force fields \cite{behler2007generalized,schutt2018schnet,wang2018deepmd, bartok2010gaussian, thompson2015spectral,gilmer2017neural}.
In the first phase of the RBMD, we implement the AMBER and CHARMM force fields, which are often used for studying the structural and dynamic properties of organic molecules and biomolecules. 

In classical force fields, the potential energy is typically separated into bonded and nonbonded potentials. The bonded potential is composed of the contributions of chemical bonds, bond angles, dihedral angles and improper dihedrals, respectively, 
\begin{equation} 
V_{\text {b}}=\sum_{\text {bonds }} k_{\mathrm{b}}\left(b-b_0\right)^2+\sum_{\text {angles }} k_\theta\left(\theta-\theta_0\right)^2 +\sum_{\text {dihedral}} k_\phi\left[1+\cos \left(n \phi-\phi_0\right)\right] +\sum_{\text {imp}} k_\omega\left(\omega-\omega_0\right)^2
\label{bonded}
\end{equation}
where  $k_b, k_{\theta}, k_{\phi}$ and $k_{\omega}$ are the potential coefficients of the four contributions, and correspondingly, $b_0, \theta_0, \phi_0$ and $\omega_0$ are the equilibrium coefficients of bond, angle, dihedral and improper energies. 
The nonbonded potential includes van der Waals and electrostatic interactions,  $V_{\text {nb}}=V_\mathrm{lj}+V_\text {elec}$, where the van der Waals interactions are characterized by the Lennard-Jones (LJ) potential 
\begin{equation}
	V_\mathrm{lj}=\sum_{i<j} 4 \epsilon_{i j}\left[\left(\frac{\sigma_{i j}}{r_{i j}}\right)^{12}-\left(\frac{\sigma_{i j}}{r_{i j}}\right)^6\right] \label{lj}
\end{equation}
and the electrostatic potential $V_\text {elec}$ reads
\begin{equation}
       V_\text{elec}=\frac{1}{4 \pi \epsilon_0}\sum_{i<j} \frac{q_i q_j}{ r_{ij}}   
\label{elec}
\end{equation}
with $\epsilon_0$ being the vacuum dielectric constant.
In Eqs. \eqref{lj} and \eqref{elec}, the summations run over all particle pairs, including interactions with image particles due to periodic boundary conditions. The coefficients for the LJ interactions are given by the Lorentz-Berthelot mixing rules \cite{frenkel2023understanding}, 
 \begin{equation} 
\varepsilon_{ij}=\sqrt{\varepsilon_i\varepsilon_j}, ~\hbox{and }\sigma_{ij}=\frac{\sigma_i+\sigma_j}{2} 
 \end{equation}
where $\varepsilon_{i}$ and $\varepsilon_{j}$ are the energy coefficients of particles $i$ and $j$, and $\sigma_{i}$ and $\sigma_{j}$ are the corresponding distance coefficients.  

\begin{figure}[t!]
\centering
\includegraphics[width=0.85\linewidth]{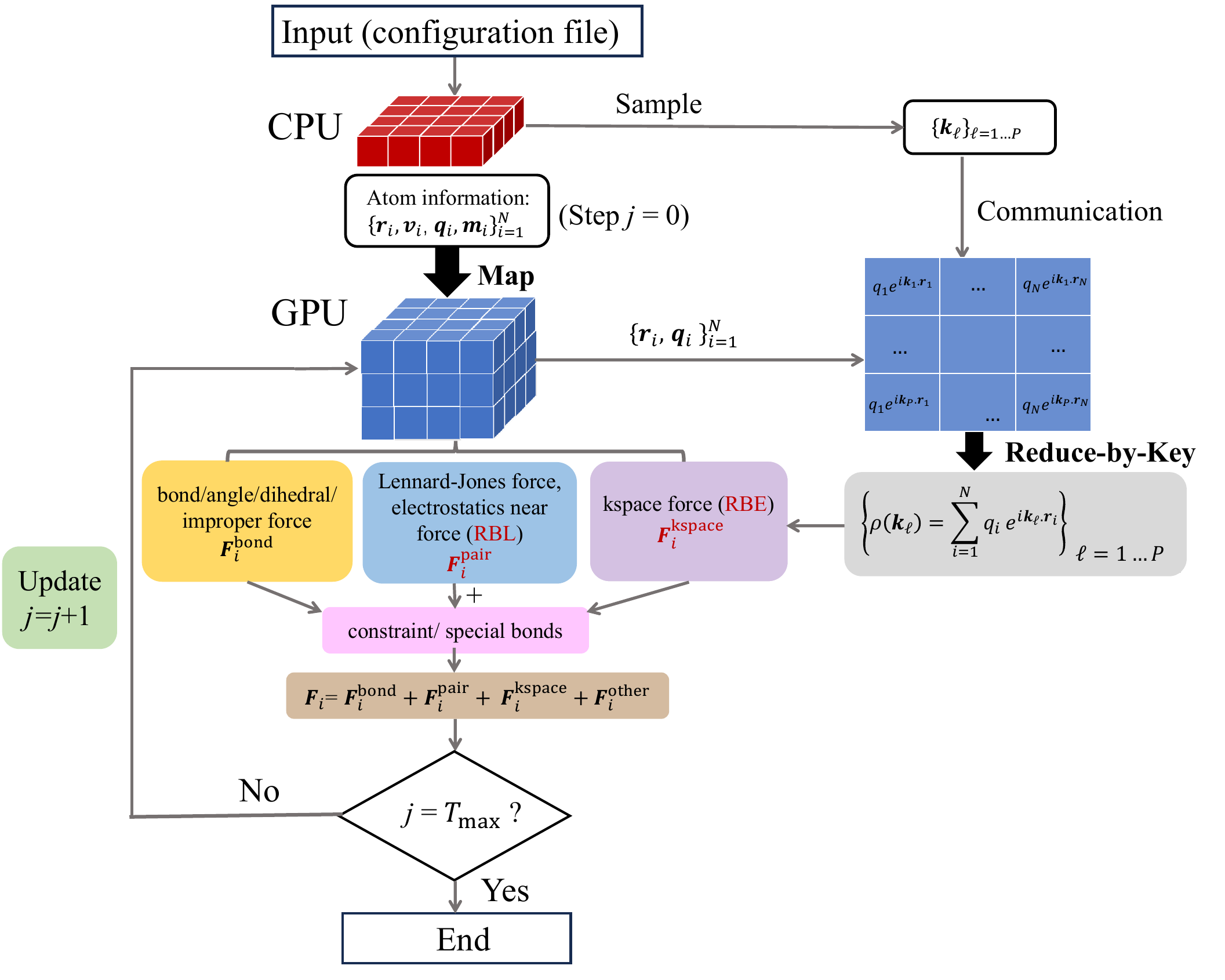}
\caption{
Simulation flowchart of RBMD. Particularly, in the RBE implementation, $P$ samples are extracted in the CPU and then transferred to the GPU. The \textit{Reduce-by-Key} operation is employed to simultaneously calculate $P$ structure factors, thereby computing the RBE force.}
\label{RBMD_workflow}
\end{figure}

The MD simulation flowchart of the RBMD is depicted in Figure \ref{RBMD_workflow}. The flowchart starts with reading the input configuration file on the CPU to collect essential particle information such as position, velocity, charge and mass. Subsequently, the particle information is transferred to the GPU (blue blocks) using the \textit{Map} operation, which will be discussed in Section \ref{vtkm_gpu}. On the GPU, the interactions of particle bonds, nonbonded short-range interactions, and long-range interactions are computed. It is noted that the nonbonded forces are the computational bottleneck of the MD. The algorithms for nonbonded forces and implementations will be discussed heavily in later sections. Given the force field, the configuration file specifies whether to apply constraints to maintain the rigid structure of certain special molecules, such as water molecules~\cite{andersen1983rattle}. For molecules with dihedral and improper angles, the RBMD implements special bonds~\cite{cornell1995second,mackerell1998all} that enable the adjustment of weights for LJ and/or electrostatic interactions between pairs of atoms that are also permanently bonded to each other. The user can configure the option "special$\_$bonds" in the configuration file. Combining all the forces computed by the above operations, the whole force on each atom is obtained. The sum of these four forces is used to update the new particle positions and velocities. Subsequently, the Velocity-Verlet algorithm \cite{frenkel2023understanding} is employed to update the new positions and velocities of particles by integrating Newton's equations of motion. If the simulation step reaches the predefined $T_{max}$ step, the iteration terminates.

The RBMD is able to compute required physical quantities, such as temperature, potential energy, radial distribution function (RDF), mean square displacement (MSD), and velocity autocorrelation function (VACF). Meanwhile, the RBMD generates the trajectory file of particles to analyze the structural properties via the VMD\cite{humphrey1996vmd} or OVITO\cite{stukowski2009visualization} packages.  Furthermore, NVE (constant volume $V$ and energy $E$) and NVT (constant volume $V$ and temperature $T$) ensemble simulations are supported on RBMD. To sample the correct NVT ensemble, the RBMD implements several thermostats, including Berendsen~\cite{berendsen1984molecular}, Langevin~\cite{schneider1978molecular} and Nos\'e-Hoover thermostats~\cite{evans1985nose}. Moreover, the RBMD realizes the module for the NVE, which is based on an energy stable scheme via an energy bath \cite{liang2024energy}.  
Based on the description of the overall workflow of the RBMD, we design the corresponding configuration file,  which is shown in the README at {\em https://github.com/randbatch-md/rbmd}.

\subsection{Electrostatic interactions}\label{rbe_rbl}

Electrostatic interactions are generally considered as computational bottleneck in MD simulations\cite{rodgers2008local,walker2011electrostatics,hu2014infinite}. To address this challenge, a common approach is to apply Ewald splitting \cite{Ewald1921AnnPhys} to separate the Coulomb kernel $1/r$ into rapidly decaying short-range and smooth long-range parts and then efficiently solve them in real space and Fourier space, respectively. This results in a series of lattice-based Ewald algorithms \cite{Hockney1988Computer, essmann1995smooth, darden1993particle, deserno1998mesh,dos2016simulations} with the FFT acceleration.
 However, the FFT is communication-intensive, requiring significant communication overhead in multi-CPU parallel computing, posing a scalability bottleneck. The RBE method overcomes this simulation bottleneck due to the use of small but mini-batches at each step in approximating the force \cite{jin2021random}, and this section gives an overview of this algorithm. 

Without loss of generality, consider $N$ charged particles in a cubic box of length $L$ with the periodic boundary condition. Denote the charge and position of the $i$th particle by $q_i$ and $\bm{r}_i$, the volume of the box by $V=L^3$. Assuming that the system is charge-neutral, $\sum_i q_i=0$. Then the total electrostatic energy of this system is given by
\begin{equation} \label{energy}
	V_{elec}= \frac{1}{2} \sum_{i,j=1}^N {\sum_{\bm{n}}}'
  \frac{q_i q_j}{|\bm{r}_{i j}+\bm{n} L|}  
\end{equation}
where $\bm{n}$ runs over the three-dimensional integer vectors, and the prime after the second sum means that $\bm{n} = 0$ is not included when $i = j$. Eq. \eqref{energy} is conditionally convergent, and one cannot obtain its approximation by a direct truncation. 

To sum up the series in Eq. \eqref{energy}, the RBE also starts with the kernel splitting  similar to the lattice-based Ewald-type algorithms,  
\begin{equation}
	\frac{1}{r}=\frac{\operatorname{erfc}(\sqrt{\alpha} r)}{r}+\frac{\hbox{erf}(\sqrt{\alpha} r)}{r}
	\label{erfcr}
\end{equation}
where $\hbox{erf}(x)$ is the error function  
\begin{equation} 
\operatorname{erf}(x)=\frac{2}{\sqrt{\pi}} \int_0^x \exp \left(-u^2\right) d u   
\end{equation}
and $\operatorname{erfc}(x)=1-\operatorname{erf}(x)$ is the error complementary function. The positive parameter $\alpha$ is determined by the requirement for efficiency.  The second term in \eqref{erfcr} is a long-range part, which is  smooth without singularity and can be transformed into a Fourier series of rapid decay. If one performs the Fourier expansion for the pairwise sum of the $\hbox{erf}(\sqrt{\alpha} r)/r$ kernel, the electrostatic energy can be rewritten into   
\begin{equation} 
	V_{elec}= \frac{1}{2}  \sum_{i,j=1}^N {\sum_{\bm{n}}}'
 q_i q_j \frac{\operatorname{erfc}\left(\sqrt{\alpha}\left|\boldsymbol{r}_{i j}+\boldsymbol{n} L\right|\right)}{\left|\bm{r}_{i j}+\boldsymbol{n} L\right|} +  \frac{2 \pi}{V} \sum_{\bm{k}\neq 0} \frac{1}{k^2}|\rho(\bm{k})|^2 e^{-k^2 / 4 \alpha}-\sqrt{\frac{\alpha}{\pi}} \sum_{i=1}^N q_i^2  
\end{equation}
where the Fourier grid $\boldsymbol{k}=2\pi\boldsymbol{m}/L$ with $\boldsymbol{m}\in \mathbb{Z}^3$. Here, the three terms correspond to the short-range, the long-range and the self-energy contributions, respectively. And
\begin{equation}
	\rho(\boldsymbol{k})=\sum_{i=1}^N q_ie^{i\boldsymbol{k}\cdot\boldsymbol{r}_i}
\end{equation}
is the structure factor of a given particle configuration. 
Then the force on the $i$th particle is given by the negative gradient of $V_{elec}$ with respect to $\bm{r}_i$, expressed by $\bm{F}_i=\bm{F}^{short}_i+\bm{F}^{long}_i$, where
\begin{equation} 
		\bm{F}^{short}_i = -q_i {\sum_{j, \bm{n}}}' q_j \left( \frac{\operatorname{erfc}(\sqrt{\alpha} r)}{r^2}+\frac{2 \sqrt{\alpha} e^{-\alpha r^2}}{\sqrt{\pi} r}\right) \frac{\bm{r}_{i j}+\bm{n} L}{\left|\bm{r}_{i j}+\bm{n} L\right|} 
	\label{fshort}
\end{equation}  
\begin{equation} 
		\bm{F}^{long}_i = -\sum_{\bm{k} \neq 0} \frac{4 \pi q_i \bm{k}}{V k^2} e^{-k^2 /4\alpha} \operatorname{Im}\left(e^{-i \bm{k} \cdot \bm{r}_i} \rho(\bm{k})\right). 
	\label{long}
\end{equation}
The short-range term Eq. \eqref{fshort} involves the error complementary and exponential functions, and both of them converge rapidly in real space.  It is feasible to compute this part of the force by determining a cutoff radius and calculating the forces exerted by particles in the sphere of radius $r_s$ centered on particle $i$, which will be computed along with other short-range forces such as LJ forces. 
The algorithm for this contribution will be discussed in Section \ref{sec::short}.

The RBE is an approach to treating the long-range part, which is expanded into a Fourier series in Eq. \eqref{long}. It avoids using the communication-intensive FFT. Instead, it approximates the Fourier series via importance sampling. In  Eq. \eqref{long}, the series can be considered as the expectation of a random variable with probability distribution $\exp(-k^2/4\alpha)/Q$ where $Q$ is the normalization constant. One can simply obtain $P$ samples, $\{\bm{k}_\ell\}_{\ell=1}^{P}$ from this probability density function, such as via the Metropolis Monte Carlo, and then an approximate expression of the long-range force can be obtained,
\begin{equation}
\bm{F}^{long}_i \approx -\sum_{\ell=1}^P\frac{Q}{P}\frac{4\pi q_i\boldsymbol{k}_{\ell}}{V k_{\ell}^2}\text{Im}(e^{-i\boldsymbol{k}_{\ell}\cdot \boldsymbol{r}_i}\rho(\boldsymbol{k}_{\ell})).
\label{frbe}
\end{equation}
Under the assumption that the samples are independent and identically distributed, one can prove that the estimate \eqref{frbe} is unbiased. Furthermore, with the Debye-H\"uckel assumption, the variance of the force estimate is proportional to the product of $(N/V)^{4/3}$ and $1/P$, and the RBE algorithm is expected to work well for an appropriate $P$ when the particle density is not large.

It has been demonstrated in \cite{jin2021random} that, when the system density is fixed, the required batch size $P$ for the same variance is independent of the number of particles $N$. As a result, the computational complexity for calculating the long-range force is $O(N)$.
From another point of view, the key issue affecting parallel efficiency for RBE is the calculation of $P$ terms of $\rho(\boldsymbol{k})$, which fits the thousands of threads of GPU simultaneous computation. Hence, the acceleration by GPU for RBE is huge compared with the CPU version of RBE expectedly. Meanwhile, the RBE avoids the mesh mapping process used in PPPM method on CPU and executes the main computation (except for the sampling procedure) on GPU, greatly  showing the advantage of RBE method.

There are many further developments after the work of Jin {\it et al.}\cite{jin2021random}. The superscalabity of the RBE for parallel computing has been demonstrated in \cite{liang2022superscalability}. Moreover, the RBE is extended to simulate systems at the NPT and NVE ensembles \cite{liang2022random,liang2024energy}, successfully capturing the long-range electrostatic correlations\cite{hu2022symmetry,gao2023screening}. It is remarked that the sum-of-Gaussians decomposition is also used to split the Coulomb kernel into near and far parts. This leads to the u-series method \cite{predescu2020u,shaw2021anton} and the random batch sum-of-Gaussians (RBSOG) method \cite{liang2023random} for electrostatic forces. The RBSOG inherits the superscalability of RBE and offers greater flexibility for extending to general kernels. It will be integrated into future developments of the RBMD.

\subsection{Short range interactions}
\label{sec::short}
Computing short-range interactions is another fundamental yet time-consuming task in MD simulations. Various techniques have been extensively discussed in the literature\cite{plimpton1995fast}. The Verlet list~\cite{PhysRev.159.98} and linked cell list~\cite{yao2004improved} algorithms, among the earliest in the field, are pivotal to early MD software packages. Their ideas are straightforward, involving constructing a stencil of bins to identify possible neighbors, binning atoms, and looping through the stencil to assemble the neighbor list. Despite achieving linear complexity, these methods require substantial memory resources and intensive processor-memory communication, limiting the accessible scale of simulations. Recent progress has explored single-instruction multiple data (SIMD) parallelization~\cite{pall2013flexible,gromacsJCP2020}. The resultant cluster pair algorithm delivers excellent performance, yet it may introduce artificial effects for the dynamics~\cite{kim2023neighbor}, and the issue of high memory consumption persists.

The RBL~\cite{liang2021random} is a recently developed stochastic method designed to accelerate short-range calculations and reduce memory usage. Unlike previous methods, the RBL introduces an outer cutoff $r_s$ to truncate the potential and an additional inner cutoff $r_c<r_s$, constructing two-level core-shell structured neighbor lists around each particle. Let 
\begin{equation}
\mathcal{C}(i):=\{j\neq i: r_{ij}<r_c\}\quad \text{and}\quad \mathcal{S}(i)=\{j:r_c\leq r_{ij}\leq r_s\} 
\end{equation}
be the neighbor sets of the $i$th particle in the core and shell regions, respectively. To demonstrate how the RBL method works, one first  decomposes the short-range force into the contributions from the two regions, 
\begin{equation}
\bm{F}_{i}^{short}=\bm{F}_i^{core}+\bm{F}_i^{shell}, \label{short}
\end{equation}
where $\bm{F}_i^{core}$ and $\bm{F}_i^{shell}$ include the interactions with all particles in $\mathcal{C}(i)$ and $\mathcal{S}(i)$, respectively. 
Direct summation is used for $\bm{F}_{i}^{core}$. Due to the singular kernel of the LJ interaction,  $\bm{F}_i^{core}$ is directly summed. For the core contribution, the RBL proposes a stochastic approximation. One randomly chooses a few $p$  particles from $\mathcal{S}(i)$ into a batch $\mathcal{B}(i)$, ignoring other neighbors in the zone. Let $N_{\text{s}}$ be the number of particles in $\mathcal{S}(i)$. Following the random batch idea~\cite{jin2020random}, one has the following unbiased estimator for the shell contribution,
\begin{equation}
\bm{F}_i^{shell*} = \frac{N_{\text{s}}}{p}\sum_{j\in\mathcal{B}(i)}\bm{F}_{ij} \label{fshell}
\end{equation}
where $\bm{F}_{ij}$ is the force due to particle $j$. 
When Eq. \eqref{fshell} replaces  $\bm{F}_i^{shell}$ in Eq. \eqref{short}, one obtains a faster algorithm for short-range interactions due to the smaller number of neighbors, which also significantly reduces the memory cost. 
One then obtains a correction to all particles, expressed by
\begin{equation}
\bm{F}^{corr} = - \frac{1}{N}\sum_{i=1}^N (\bm{F}_i^{core}+\bm{F}_i^{shell*})  
\end{equation}
The forces for all particles are added by this correction force to achieve the conservation of the total momentum. Since only the interacting neighbors within the core list and the batch are stored, the RBL reduces memory usage and computational cost by a factor of $4 \pi n r_s^3/(4 \pi n r_c^3+3p)$~\cite{liang2021random} where $n=N/V$ is the particle number density. It can be demonstrated that the estimation of forces is unbiased. As the short-range force includes the LJ force and the short-range component of electrostatic force Eq.\eqref{fshort}, the variance of the short-range force estimation is in the order of $(N_s-p)n (r_c^{-11}+\text{erfc}(\sqrt{2\alpha} r_c))/p$ under the condition of a homogeneous assumption \cite{irbe2022jpca}. Consequently, the variance of the RBL algorithm decreases rapidly with the increase of $r_c$ when the density is not large, thus ensuring the accuracy of the computation.

The RBMD software integrates the RBL method to accelerate short-range interaction calculations, with the specific heterogeneous implementation detailed in Section~\ref{rbl_gpu}.

\section{Implementation details}\label{gpu_imple}
In this section, we give a detailed interpretation of GPU programming, especially the RBE and RBL algorithms, by GPU-CPU heterogeneous computing.

\subsection{VTK-m framework}\label{vtkm_gpu}

The optimized strategy of the RBE used in the previous work \cite{liang2022superscalability} depends on the AVX-512 vectorization and the MPI parallelization. Some AMD or Kunpeng CPUs rather than Intel CPUs do not support the AVX-512 instruction set, therefore, one has to design specific RBE codes for different CPUs, and this affects the scalability of the RBE algorithm. In the RBMD, the framework of visualization ToolKit for massively 
threaded architectures (VTK-m) \cite{bolstad2023vtk}, which is a wrapped third-party library to fit in different kinds of GPUs such as NVIDIA, AMD and Intel, is adopted to develop a generally usable code.

\begin{figure}[t!]
\centering
\includegraphics[width=0.85\linewidth]{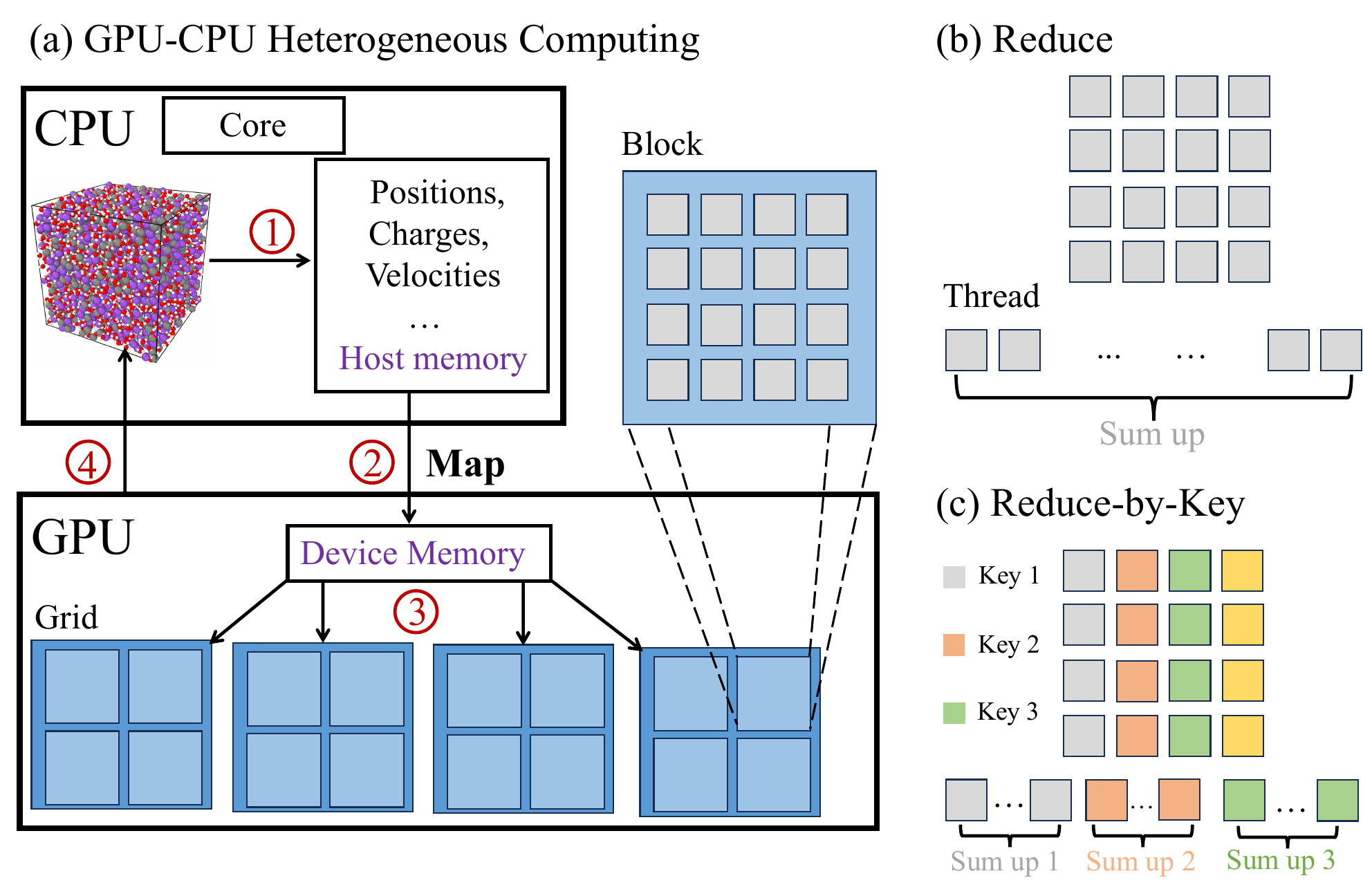}
\caption{
(a) The specific process of CPU-GPU heterogeneous computing for the RBMD. The number of steps is marked in red.  \textit{Map} is the parallel operator. The dark blue square represents "grid", the light blue square is "block" and the grey square is "thread" in GPU; (b) the parallel operator \textit{Reduce}; and (c) the parallel operator \textit{Reduce-by-Key}.
}
\label{loop_operator}
\end{figure}

The VTK-m is a framework specifically designed to accommodate different processor architectures\cite{moreland2016vtk}. It provides abstract models for data and execution, enabling their application across a variety of GPU architectures. To illustrate how the RBMD integrates with the VTK-m, Figure \ref{loop_operator}(a) schematically displays the structure of the CPU-GPU heterogeneous computing and the data transfer process between GPU and CPU. Both the computing engines include computational modules (cores or grids) and storage modules by memory. 
The difference lies in that the GPU has more computing units compared to the CPU, which allows the GPU to process large amounts of data in parallel.
When the data from the input file is read into the CPU, it is stored in the host memory (Step 1 in Figure \ref{loop_operator}(a)). Next, in Step 2, this data is transferred from the host memory to the device memory. The data in the device memory is then distributed to different grids for parallel computation (Step 3). Finally, in Step 4, once the simulation step reaches the point of thermodynamic output, the results computed by the GPU are transferred back to the CPU. This completes the heterogeneous computing operation between the GPU and CPU. It is noteworthy that the third step is the core step for the GPU computation. The VTK-m automatically distributes the tasks to different threads for parallel computation. Simultaneously, after the fourth step is completed, the GPU memory is automatically released, and the results are transferred to the CPU. 

We now introduce some details of  Step 3 of Figure \ref{loop_operator}(a), namely, how to utilize parallel operators in the VTK-m structure to accelerate the computation.
In the VTK-m structure, three parallel operators for GPU parallel acceleration: \textit{Map}, \textit{Reduce} and \textit{Reduce-by-Key} play crucial roles in implementing the RBE/RBL methods by the GPU. Firstly, the type definition for information about particles such as positions, charges and velocities is the  \textit{Map} operation. By defining the data type, the information of each particle is transferred to the device memory on the GPU, as shown in Figure \ref{loop_operator}(a). And then the parallel calculations on different threads on the GPU are executed. Secondly, the \textit{Reduce} operation, shown in Figure \ref{loop_operator}(b), allows for the parallel computation of summating all elements in an array, making it highly effective for large-scale summations. Thirdly, the \textit{Reduce-by-Key} operation in Figure \ref{loop_operator}(c) is to summate elements on different threads with the same \textit{key} in parallel, where the \textit{key} refers to an identifier of the thread. In the RBE, calculating the $P$ structure factors is the most important part for the improvement of the RBE efficiency. The operation \textit{Reduce-by-Key} can sum up $\rho(\boldsymbol{k_{\ell}})$ with the same $\{\boldsymbol{k}_{\ell}\}_{\ell=1}^P$, i.e., $\boldsymbol{k_{\ell}}$ represents the \textit{key} in Figure \ref{loop_operator}(c), in parallel. By using \textit{Reduce-by-Key}, all 
$P$ structure factors can be calculated simultaneously on the GPU, significantly improving the utilization rate of GPU memory and computational units compared to reducing them individually. Therefore, \textit{Reduce-by-Key} is essential for the acceleration of the RBE algorithm.


\subsection{Implementation of the RBE}\label{rbe_gpu}

Algorithm \ref{algo_rbe} presents the procedure for computing the RBE force. 
At the preparation stage, the \textit{Map} operator is used to allocate the position and charge  of each particle  to the GPU. The splitting parameter $\alpha$ and sample number $P$ are read from the input file on the CPU. The RBMD first obtains $P$ samples $\{\boldsymbol{k}_{\ell}\}_{\ell=1}^P$ from the distribution $Q^{-1}e^{-|\boldsymbol{k}|^2/(4\alpha)}$ by the Metropolis  algorithm\cite{frenkel2023understanding} on the CPU, where $Q$ is the normalizing factor. In general, $P$ is far smaller than the particle number $N$. Thus, this sampling procedure takes little time on the CPU. The $P$ samples and the positions and charges from the \textit{Map} operator, are then used to compute $q_ie^{i\boldsymbol{k}_{\ell}\boldsymbol{r}_i}$ with $\ell=1,\dots, P$ on the GPU for all particles. The next step is then to use the \textit{Reduce-by-Key} operator to get $\rho(\boldsymbol{k}_{\ell})(\ell=1,\dots,P)$ by summating the results of these Fourier modes.  Finally, the RBE force Eq. (\ref{frbe}) is calculated on the GPU using structure factors $\rho(\boldsymbol{k}_{\ell})(\ell=1,\dots,P)$.

\begin{algorithm}[h] \label{algo_rbe}
\caption{Random batch Ewald implementation of the RBMD}
\KwIn{Structure: \textit{Map} each $\boldsymbol{r}_i, q_i$ of particle $i$ to the GPU \\
	Parameter: Read parameters $\alpha $ and $P$ on the CPU host}
\BlankLine
Sample sufficient number of $\boldsymbol{k}\sim Q^{-1}e^{-|\boldsymbol{k}|^2/(4\alpha)}$ with $\boldsymbol{k}\neq \boldsymbol{0}$ by the Metropolis procedure,  pick $P$ frequencies $\{\boldsymbol{k}_{\ell}\}$  and send them to the GPU \\
Compute $q_ie^{i\boldsymbol{k}_{\ell}\boldsymbol{r}_i}$ for $\ell=1,\dots, P$ on the GPU\\
\textit{Reduce-by-Key} the results of Step 2 on the GPU $\to$ $\rho(\boldsymbol{k}_{\ell}) (\ell = 1,\dots,P)$ \\
Using $\{\boldsymbol{k}_{\ell}\}$  and $\{\rho(\boldsymbol{k}_{\ell})\}$ to calculate the force Eq.~\eqref{frbe} on the GPU 
\end{algorithm}

\subsection{Implementation of the RBL}\label{rbl_gpu}
The implementation of the RBL algorithm on the GPU-CPU architecture is given in Algorithm \ref{algo_rbl}. The preparation stage mirrors that of the RBE method. The \textit{Map} operator allocates the position and charge of each particle to the GPU. Simultaneously, the RBMD takes the core radius $r_c$ and sample number $p$ on the CPU. The cell list is constructed using cubic cells with a side length of $r_c$. Subsequently, all particles within the core area of particle $i$ and $p$ randomly selected particles from its shell area are grouped to obtain the neighbor list for particle 
$i$ on the GPU. At Step 2, $\bm{F}^{core}_i$ and $\bm{F}^{shell}_i$ are computed 
on the GPU, respectively. The $Reduce$ operator then combines $\bm{F}^{core}_i$ and $\bm{F}^{shell}_i$ to obtain $\bm{F}^{corr}$. At the final step, the overall force $\bm{F}^{short}$ is calculated by summing up $\bm{F}^{core}_i$, $\bm{F}^{shell}_i$ and $\bm{F}^{corr}$ on the GPU. It should be noted that the RBMD calculates all short-range components using this RBL algorithm based on the GPU-CPU architecture, including the Lennard-Jones force and the short-range component of the electrostatic force. 

The computation of the short-range part of the electrostatic force involves an expensive error function and an exponential term in Eq.\eqref{fshort}, which significantly impacts computational efficiency. To address this, we integrate the look-up table method\cite{liang2023random} with the RBL implementation to speed up the calculation of the short-range electrostatic force. The look-up table method divides the cutoff region into two segments by introducing a $r_0<r_s$. When $r<r_0$, the force is calculated using a series of Taylor expansion. For the case of $r_0<r<r_s$, the values of the Gauss error function and exponential terms are pre-computed and stored in a table file. The force in the shell region is then approximated using polynomial interpolation from these precomputed values\cite{wolff1999tabulated}. The computational cost of the short-range component of electrostatic forces can be significantly reduced by avoiding the direct calculation of these Gaussian error and exponential functions.

\begin{algorithm}[h]\label{algo_rbl}
\caption{Random batch list implementation of the RBMD}
\KwIn{Structure: Obtain each $\boldsymbol{r}_i$ and $q_i$ of particle $i$ on the GPU\\
	Parameter: Read parameters $r_c$, $r_s$  and the sample number $p$ from the CPU and construct the cell list according to $r_c$ on the GPU}
\BlankLine 
Choose $p$ particles in the shell region randomly and all particles in the core region to build the neighbour list on the GPU \\
Compute $\boldsymbol{F}^{core}_i$ from the core region and $\boldsymbol{F}^{shell}_i$ from the shell region for each particle on the GPU \\ 
\textit{Reduce} the results of Step 2 on the GPU to get the $\boldsymbol{F}^{corr}$ \\
Compute $\boldsymbol{F}^{short}$ from Steps 2 and 3 on the GPU
\end{algorithm}

\section{Results and discussion}\label{discuss}

In this section, numerical results are presented to validate the correctness of the RBMD software and the efficiency of the random batch methods implemented in machines with GPU-CPU heterogeneous architectures.
We perform simulations for three benchmark systems. The first system is the LJ fluid, which is a typical short-range system with the force field having only the nonbonded LJ potential. The second system is a $1:1$ electrolyte system based on the primitive model, where the force field includes both the LJ and electrostatic interactions \cite{liang2022hsma}. The third system is an all-atom pure water system, utilizing the SPC/E pure water model\cite{mark2001structure}. The water model includes not only the nonbonded interactions between atoms, but also the bond and angle interactions in Eq. (\ref{bonded}). The SHAKE algorithm\cite{ryckaert1977numerical} will be used to constrain the bond and angle forces between atoms, maintaining the rigidity of the water molecules. All test systems are maintained in the NVT ensemble using the Berendsen thermostat under the periodic boundary condition in three directions.

The results are performed on three different architectures, each including one CPU core and one GPU card.  The testing GPU hardware products are Tesla V100, GeForce RTX 4090 and Tesla A100, respectively, and their parameter details on CUDA cores, memory and memory bandwidth are presented in Table \ref{gpu_settings}. The corresponding CPUs for the three hardwards are 
Intel (R) Xeon (R) Platinum 8255C CPU@2.5GHz; AMD EPYC 7402@2.8GHz; and Intel (R) Xeon (R) Gold 6230 CPU@2.1GHz.

\begin{table}[htbp]
\caption{Comparison of three evaluated NVIDIA GPUs}
\vspace{1.0em}
\centering
\begin{tabular}{lccc}	
	\toprule[1.5pt] 	
	\vspace{0.3em}
	Performance index \ GPU & Tesla V100 & GeForce RTX 4090 & Tesla A100 \\
	
	
	\midrule[1pt]

    Release Year & 2018 & 2022 & 2020 \\
    Architecture & Volta & Ada Lovelace & Ampere \\
    Process Size & 12 nm & 5 nm & 7 nm\\
	CUDA Cores & 5120 & 16384 & 6912\\
    Memory Type & HBM2 & GDDR6X & HBM2e \\
	GPU memory (GB) & 32 &  24 & 40    \\
	Memory bandwidth (GB/s) & 900 & 1008 &1555     \\
	
	\bottomrule[1.5pt]
\end{tabular}
\label{gpu_settings}
\end{table}

\subsection{Accuracy} \label{accuracy}

We first test the accuracy of RBMD on three benchmark systems, performed on the RTX 4090 GPU card.
For the LJ liquid, the length of the computational box in each direction is $L=10\sigma$, where $\sigma$ is the reduced length unit\cite{Allen2017ComputerLiquids}, with  $N=1000$ atoms in the system. The cutoff radius takes $r_s=5\sigma$, and the core radius for the RBL method sets $r_c=2.5\sigma$. The time step takes 0.002$\tau$ with $\tau=\sigma\sqrt{m_0/(k_BT)}$ being the unit of time and $m_0 = 1$ (LJ unit) the ion mass. The electrolyte solution system uses the same $L$, $r_s$, $r_c$ and the time step as the LJ system, but with 500 cations and anions. In the dimensional unit, the electrostatic potential between particles is given by $V_{ij}=\ell_B q_iq_j/r_{ij}$ where the Bjerrum length $\ell_B = 3.5\sigma$. In the RBE, $\alpha = 0.362$, which is consistent with the choice of the PPPM method. In both systems, the equilibrium temperature is set $k_BT =1$ with $k_B$ being the Boltzmann constant. The batch sizes for the RBL and RBE 
are set at $p=100$ and $P=100$, respectively. The third benchmark is an all-atom water system. The water molecule is described by the SPC/E model. The simulation box has a length of $66.9 \textup{~\AA}$ with 1000 water molecules. The SHAKE algorithm was employed to constrain the bond lengths and bond angles of water molecules. The RBE has a batch size $P=100$ and Ewald splitting parameter $\alpha = 0.048$. The short-range cutoff radius $r_s=12\textup{~\AA}$ and the RBL core radius $r_c=6\textup{~\AA}$. The RBL batch size $p=100$, the same as the other two examples. The time step for simulation is $dt=1 fs$ and the equilibrium temperature is $T=298K$.

\begin{figure}[t!]
\centering
\includegraphics[width=0.8\linewidth]{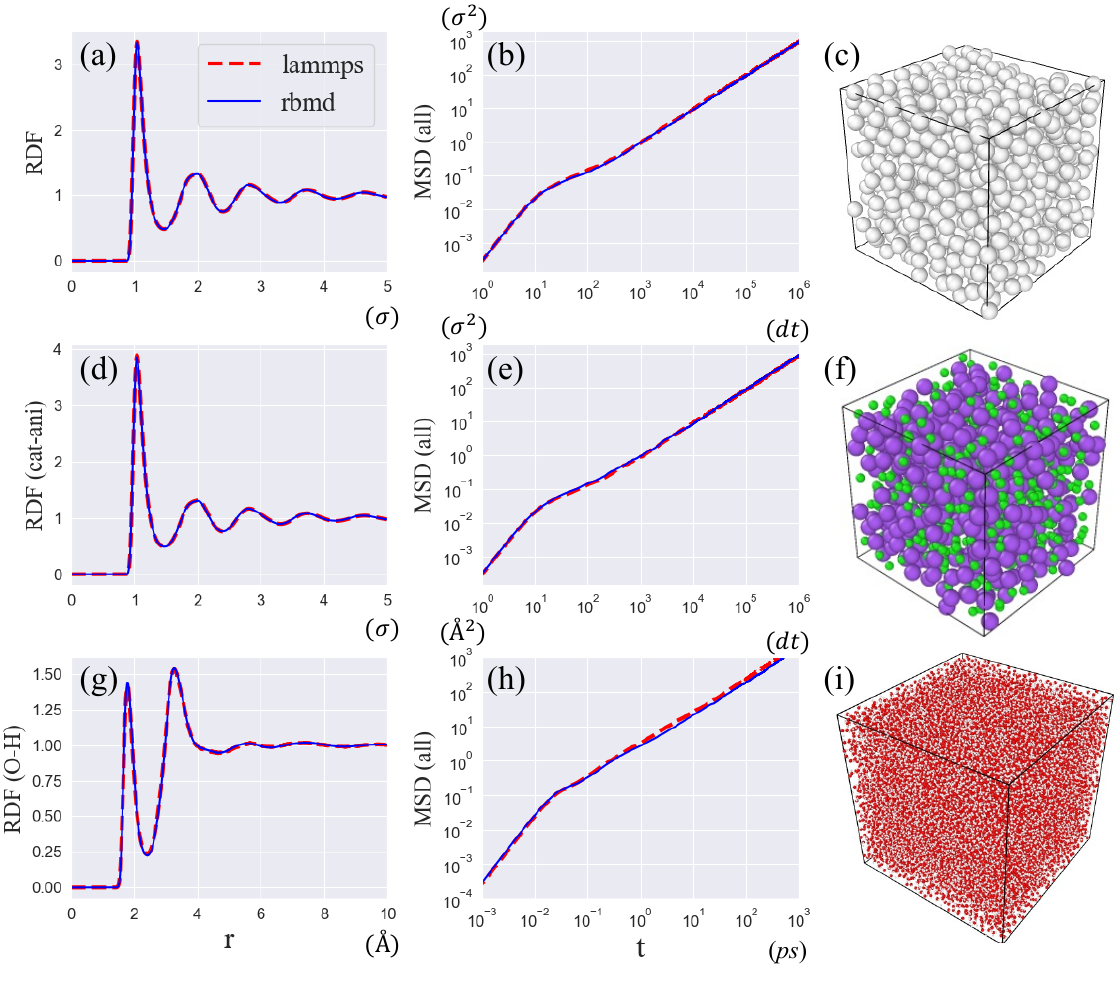}
\caption{ Results of radial distribution function (RDF),  mean squared displacement (MSD), and snapshots produced by OVITO for three benchmark systems: (a-c) the Lennard-Jones fluid; (d-f) the electrolyte solution; and (g-h) the SPC/E water system. The RBMD and LAMMPS are shown to be in good agreement.}
\label{result_rbmd}
\end{figure}

For all three examples, We run  $5\times10^5$ steps for equilibrium and $1.0\times 10^6$ steps for statistics. We calculate the RDFs and MSDs by both the RBMD and the LAMMPS, and the results are depicted in Figure \ref{result_rbmd}. These results demonstrate that the RBMD reproduces the LAMMPS results well.  
The accuracy of the RBE and RBL has been extensively reported in existing literature \cite{jin2021random,liang2021random,liang2022superscalability,liang2022random,irbe2022jpca}, and the RBMD results are also in consistent with these previous studies.

\subsection{Efficiency}
Our investigation below will focus on the performance of the RBMD, in comparison with the LAMMPS with the PPPM method for the electrostatic solver and the Verlet list for the near-field calculations.  
To evaluate the performance of LAMMPS with PPPM GPU implementation, we load the "GPU", "KSPACE" and "RIGID" packages of the Stable Release (2 Aug 2023). To enable GPU-accelerated near and far parts of the LAMMPS, the "pair$\_$style" and "kspace$\_$style" options are "lj/cut/coul/long/gpu" and "pppm/gpu" respectively, in the electrolyte solution and pure water systems. The "pair$\_$style" setup is "lj/cut/gpu" for the LJ system. And then we use the command "mpirun -n 1 lmp -sf gpu -pk gpu 1 -i in.file" to execute the LAMMPS file. The wall-clock times for long-range and short-range parts are then recorded in the "Kspace" and "Pair" subroutines. 
The experimental configuration for performance comparison between the RBMD and LAMMPS is conducted on a single GPU card with one CPU core. In the PPPM implementation of the LAMMPS, the charge interpolation and the force calculation are accelerated on the GPU (device). As FFT computation requires a large amount of MPI communications, this part is running on the CPU (host)\cite{george2022review}. Also, the transfer of particle information between the GPU (device) and CPU (host)\cite{THOMPSON2022108171} is required for each time step. Therefore, for simulations using a single GPU, the LAMMPS generally uses  multi-core CPUs to achieve high efficiency. In the RBMD, only sampling of $P$ frequencies is performed on the CPU for the RBE algorithm, and a single CPU core can efficiently handle this task. Meanwhile, the calculation of the RBL algorithm is entirely performed on the GPU. Therefore, the RBMD results in efficiency are obviously dominant, mostly because of the low efficiency of the GPU usage in the LAMMPS under the limitation of one CPU core.

\begin{figure}[t!]
\vspace{-3em}
\centering
\includegraphics[width=1.0\linewidth]{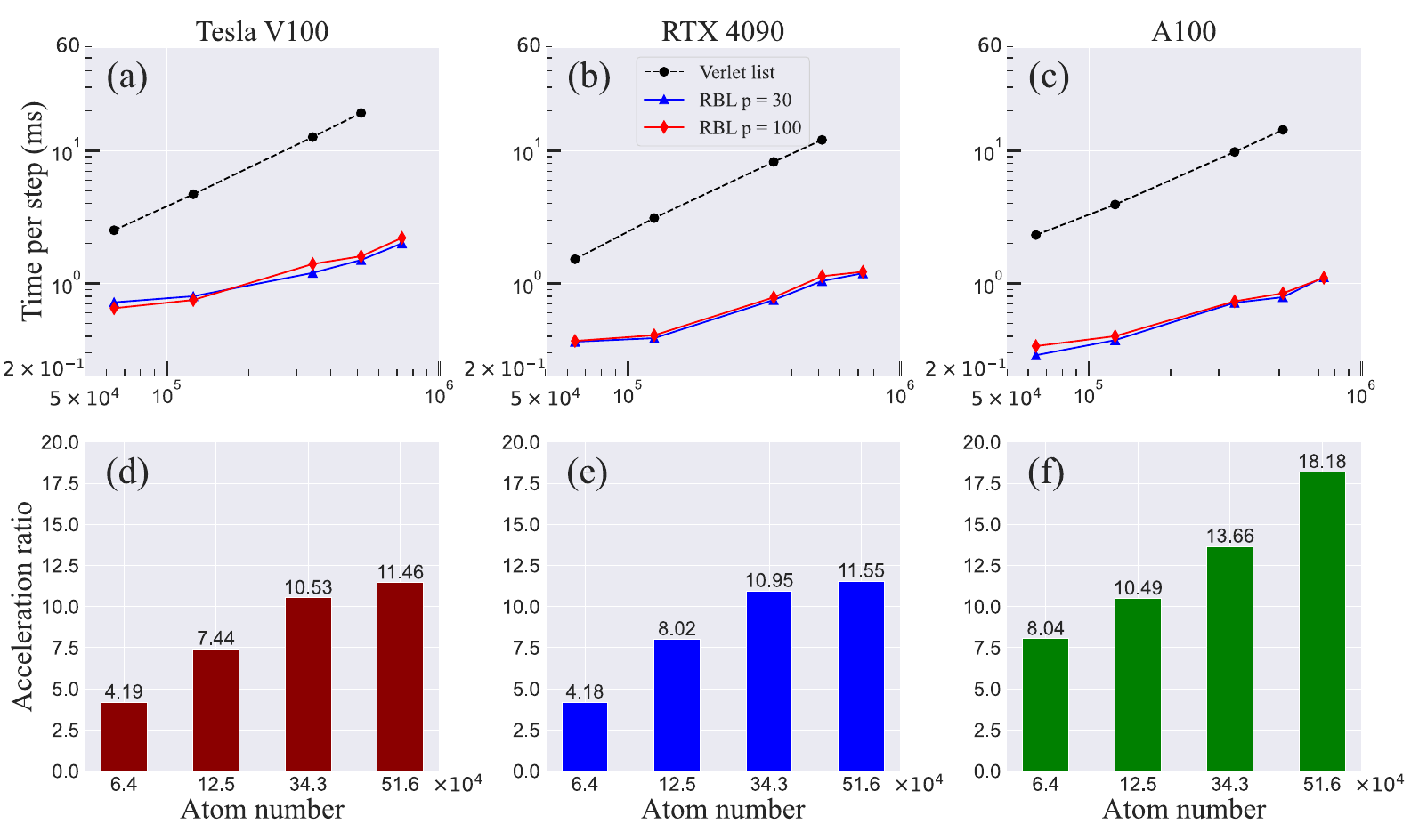}
\caption{Wall-clock time for the Lennard-Jones fluid on the Tesla V100 (a), RTX 4090 (b) and A100 (c) architectures. The results calculated by the RBL method in the RBMD and the Verlet list method are present. The results of the acceleration ratio between the RBL algorithm ($p=30$) and the Verlet list method on the Tesla V100, RTX 4090 and A100 architectures are depicted in (d-f), respectively.}
\label{GPU_RBL_per_step}
\end{figure}

{\it The Lennard-Jones fluid} ---- We investigate the efficiency of the RBMD for the LJ system.
In the calculations, the same setup as in Section \ref{accuracy} is used for the force field parameters, but with two batch sizes $p=30$ and $100$. The particle number density is fixed to be $0.01\sigma^{-3}$, the same as in \cite{irbe2022jpca}. 
Figure \ref{GPU_RBL_per_step}(a-c) shows the results of the wall-clock time per MD step in three different GPUs for the particle number $N$ in the range from $5\times 10^4$ to $10^6$, calculated via the RBL in the RBMD and the Verlet list in the LAMMPS.
It is shown that both the RBL and the Verlet list are almost linearly scaling with particle number $N$ across three GPUs. The RBL results for the two batch sizes are close. Overall, the RBL achieves  one order of magnitude faster than the direct Verlet list method on the three different GPUs. Figure \ref{GPU_RBL_per_step}(d-f) displays the acceleration ratios of 4 different system sizes for $p=30$ in the three machines.  
The most prominent speedup is on the A100 machine, which achieves a factor 12.59 faster than the reference calculation on average. For a typical system of $10^5$ particles, the wall-clock time per step for the RBL of $p=30$ is 0.35 ms,  corresponding to $2.47\times 10^8$ steps per day. These attractive performances in the wall-clock time demonstrate the efficiency of the GPU-accelerated RBL algorithm.


\begin{figure}[t!]
\centering
\includegraphics[width=1.0\linewidth]{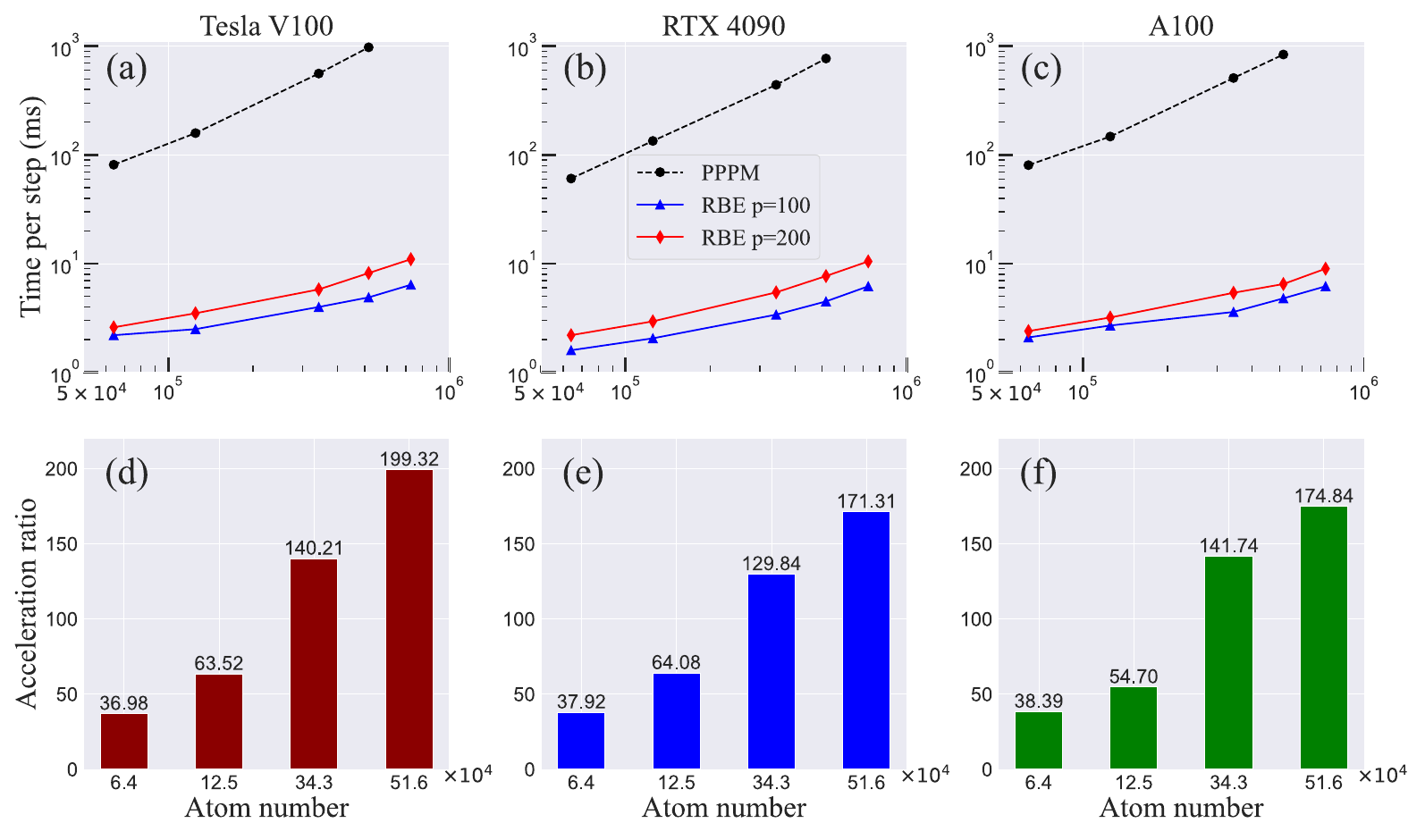}
\caption{Wall-clock time for kspace calculations of the primitive electrolyte solution on the Tesla V100 (a), RTX 4090 (b) and A100 (c) architectures. The results calculated by the RBE in the RBMD and the PPPM in the LAMMPS are present. The results of the acceleration ratio between the RBE algorithm ($P = 100$) and the PPPM method on the Tesla V100, RTX 4090 and A100 architectures are shown in (d-f), respectively.}
\label{GPU_RBE_per_step}
\end{figure}

{\it The primitive electrolyte solution} ----
The primitive electrolyte solution is used as the test case for the efficiency of the RBE implementation. The Parameters $r_s=12\sigma$ and $r_c=6\sigma$ and the particle number density are the same as for the Lennard-Jones fluid. Figure \ref{GPU_RBE_per_step}(a-c) illustrates the comparison of kspace wall-clock times using the RBE and PPPM algorithms for systems ranging from $5\times 10^4$ to $10^6$ particles on three different architectures.  Figure \ref{GPU_RBE_per_step}(d-f) presents  the acceleration ratios of 4 different system sizes. It can be seen that the computation times for both the RBE and PPPM algorithms increase linearly with the number of particles. Additionally, the RBE algorithm is 1-2 orders of magnitude faster than the PPPM algorithm across all three devices. Specially for $10^5$ number of particles on the A100, the wall-clock times per step for the RBE algorithm of $P=100$ and 200 are about 2.43 ms and 2.89 ms, whereas 127.31 ms for the PPPM algorithm. The acceleration ratio is 52.39 and 44.05, respectively, which illustrates the efficiency of the RBE run is high. One notes that the PPPM is pretty slow because only one CPU core is used. In practice, one usually runs the simulations with a combination of multicores and the GPU for better performance of the LAMMPS.

\begin{figure}[t!]
\centering
\includegraphics[width=1.0\linewidth]{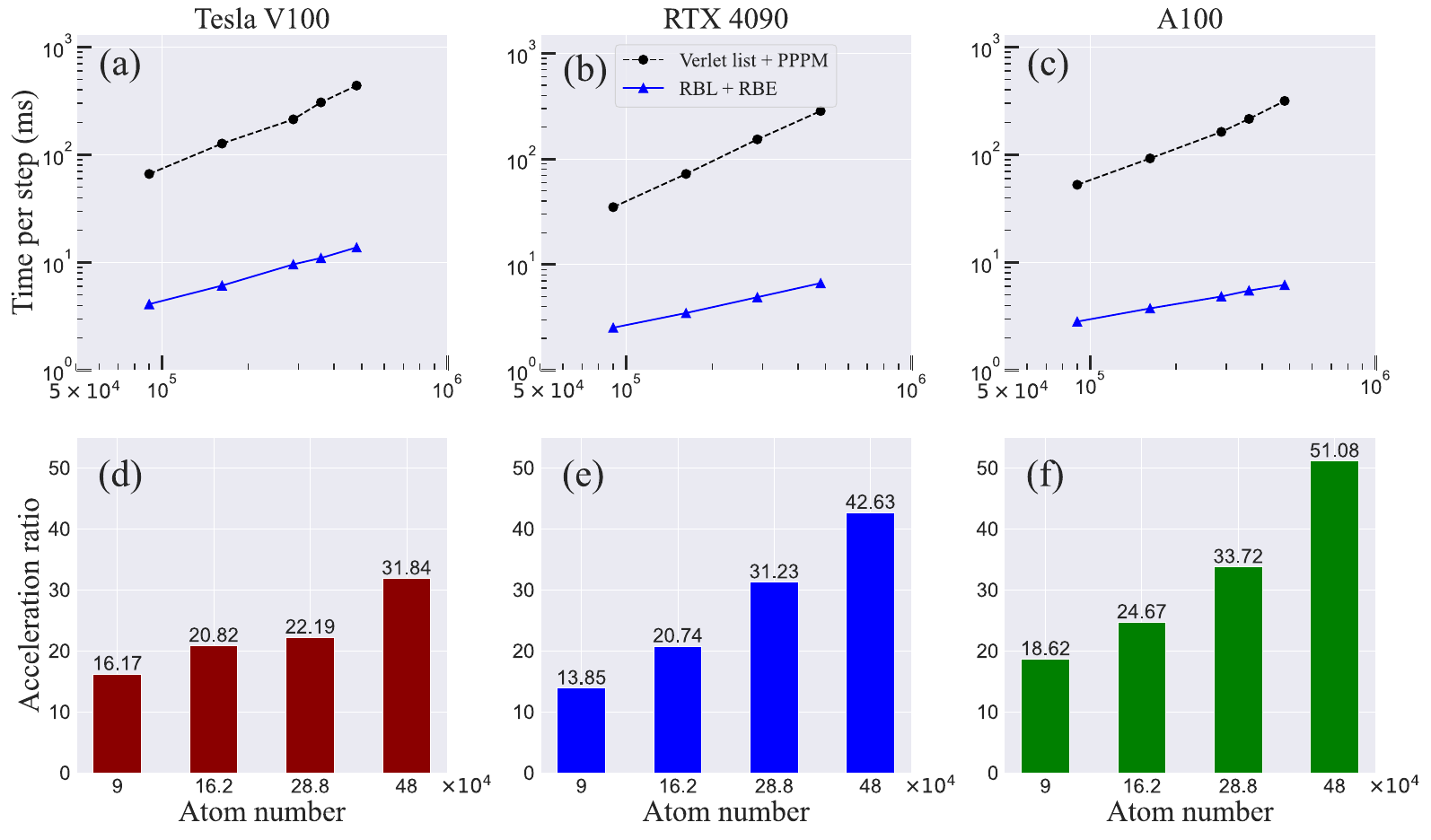}
\caption{Wall-clock time for nonbonded calculations of the all-atom water system on the Tesla V100 (a), RTX 4090 (b) and A100 (c) architectures. The results calculated by the RBL and RBE methods in the RBMD and the Verlet list and PPPM methods in the LAMMPS are present. The results of the acceleration ratio between the RBL ($p = 30$)  and RBE ($P = 100$) algorithms with the Verlet list method and the PPPM method on the Tesla V100, RTX 4090 and A100 architectures are shown in (d-f), respectively.}
\label{GPU_RBE_RBL_per_step}
\end{figure}

{\it The all-atom SPC/E water} ----
We test the performance of the RBMD for all-atom systems by using an all-atom SPC/E water system. We choose  $r_s=12\textup{~\AA}$, $r_c=6\textup{~\AA}$ and $p=30$ in the RBL algorithm and $P=100$ in the RBE algorithm for the pure water system, and the particle number density is $0.01\textup{~\AA}^{-3}$. The results are  depicted in Figure \ref{GPU_RBE_RBL_per_step}, where the linear growth with increasing atom numbers by the RBMD is observed, similar to Figures \ref{GPU_RBL_per_step} and \ref{GPU_RBE_per_step}. The RBL and RBE reach  one order of magnitude faster than the Verlet list method and the PPPM method on the three different GPUs. Figure \ref{GPU_RBE_RBL_per_step}(d-f) presents the acceleration ratios for 4 different system sizes, illustrating a $13\sim 51$ factor of speedup in comparison to the LAMMPS for the three hardware products. For the atom scale of $10^5$ on the A100, the wall-clock time per step for nonbonded computation is 3.06 ms by the RBL and RBE algorithms, 57.02 ms by Verlet list and PPPM algorithms. Thus, the acceleration ratio is 18.63, which also demonstrates the efficiency of RBE and RBL algorithms in enhancing the acceleration of nonbonded interactions for all-atom SPC/E water system by GPU-CPU heterogeneous computing.


\subsection{Large-scale systems}

\begin{figure}[t!]
\centering
\includegraphics[width=1.0\linewidth]{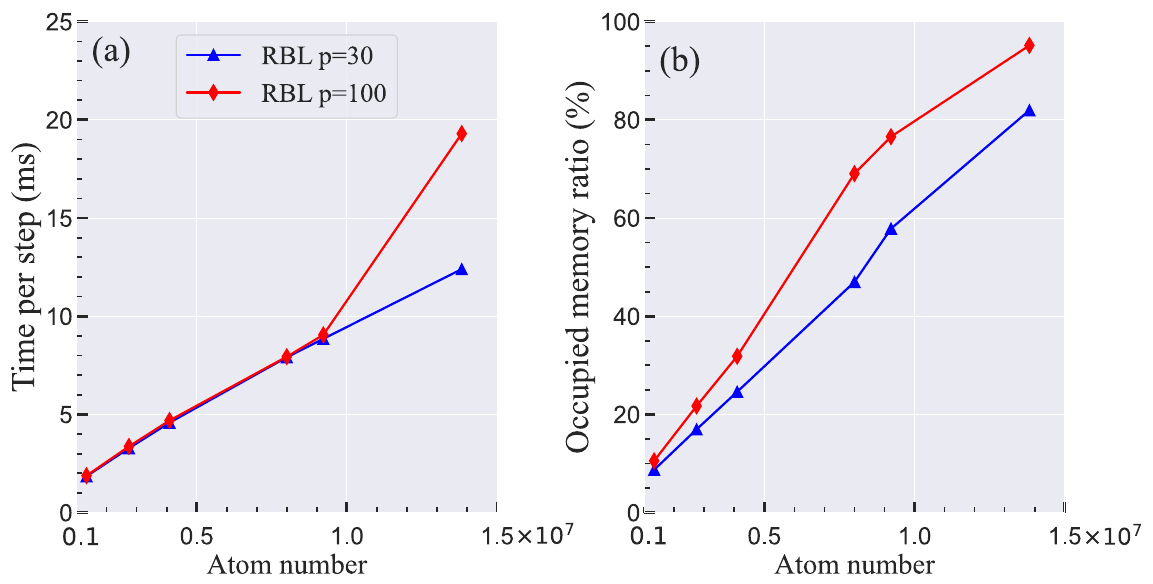}
\caption{(a) Wall-clock time per timestep (ms) on near-field computation for Lennard-Jones system by RBMD on the A100; (b) Occupied GPU memory ratio on the A100.}
\label{lj_100w_rbmd}
\end{figure}

We next demonstrate the performance of the RBMD in large-scale particle ($10^6$ to $10^7$ atom number) simulations.
First, we conduct tests on the LJ system. On the A100, we evaluate the performance for the LJ system for $p=30$ and $100$ with the particle number increasing from $1.3$ to $13.8$ million, and the results are present in Figure \ref{lj_100w_rbmd}.  One observes that as the particle number increases from 1331000 to 9216000, the time increases linearly. One can estimate that the wall-clock time for $N=10^6$ is about 1.5 ms per step, i.e., the speed is as far as 667 steps per second for computing the LJ interactions. 
When the particle number is 13824000, the memory of the GPU at $p=100$ is almost fully occupied, thus the computation speed is significantly decreased due to the communication latency. The linearity remains for $p=30$. In Figure \ref{lj_100w_rbmd}(b), the maximum occupied memory with an increase of $N$ is displayed. It is shown that the occupied GPU memory almost linearly increases with the particle numbers of the simulation box. This is due to the fact that the data of particles on the RBMD is pre-allocated on the GPU at the beginning. At $N=13824000$, the occupied GPU memory ratio at $p=100$ is $95.16\%$, which is nearly filled with the whole GPU memory of A100, approaching the von Neumann bottleneck\cite{zou2021breaking}.

\begin{figure}[t!]
\centering
\includegraphics[width=1.0\linewidth]{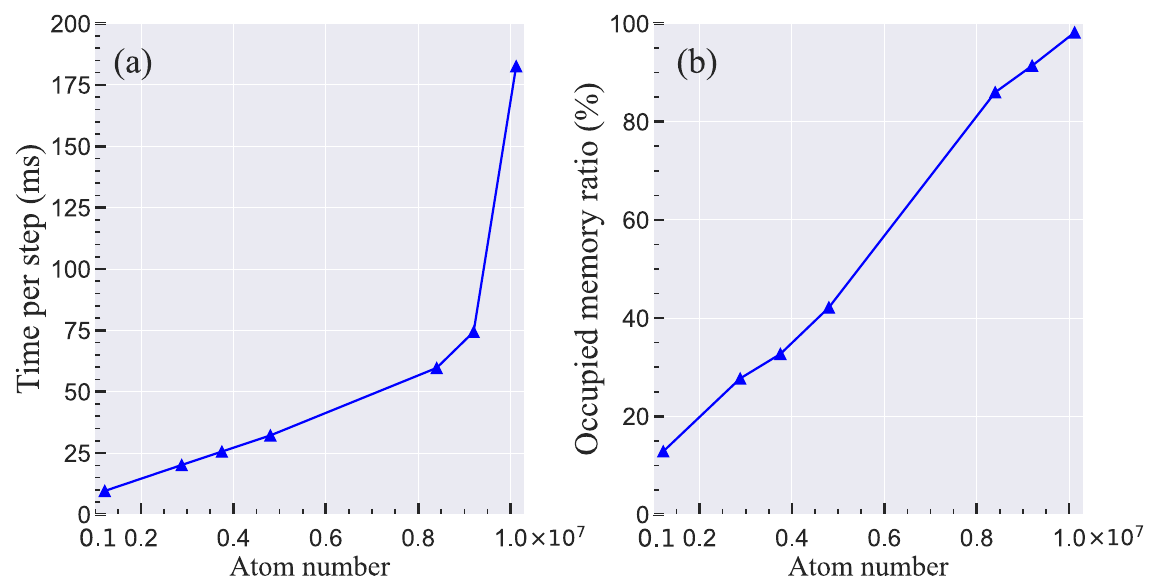}
\caption{(a) Wall-clock time per timestep (ms) for nonbonded interactions of the water system by the RBMD on the A100; (b) Occupied GPU memory ratio on the A100.}
\label{H2o_nonbond_rbmd}
\end{figure}

Figure \ref{H2o_nonbond_rbmd} presents the results for the all-atom SPC/E water system, where one takes parameters $r_s=8\textup{~\AA}$, $r_c=4\textup{~\AA}$,  $p=30$, and $P=100$. Similar results as those of the LJ system are observed.  The wall-clock time consumed by nonbonded interactions increases linearly with the number of particles, up to about $0.9\times 10^7$.  
For $N=10^6$, the wall-clock time of the RBMD for each step is approximately 8.20 ms. When the particle number reaches as high as 10,120,000, the GPU memory usage on an A100 is nearly full (98.19$\%$ in Figure \ref{H2o_nonbond_rbmd}(b)), significantly affecting the simulation efficiency due to the von Neumann bottleneck. It is noted that a smaller cutoff radius is taken to save memory, such that the RBMD can simulate a system of $10^7$ particles in a single card, which is beyond the limit of the classical simulators. One could adjust the parallel strategy of the RBE algorithm to dynamically allocate the computation of structure factors on the GPU so that the linearity can be scaled to a larger size.

\section{Concluding remarks}
\label{sec::Conclusion}

In summary, we have developed the RBMD, a novel molecular dynamics package for large-scale simulations in CPU-GPU heterogeneous architectures, which is based on the random batch methods to accelerate nonbonded interactions. 
We introduce the software workflow and present the heterogeneous acceleration of the stochastic algorithms within the VTK-m programming framework. We demonstrate the high efficiency of the RBMD for computing nonbonded interactions; typically, for $N=10^6$ particle system, the wall-clock time is at an average rate of 1.50 ms per step for Lennard-Jones fluid and 8.20 ms per step for all-atom SPC/E water. It is also shown that the RBMD allows to simulate systems of size up to ten million on a single GPU. These results highlight the unique advantages of the RBMD for large-scale computational simulations. 

Many works are to be done for the next version of the RBMD. Particularly, the random-batch sum-of-Gaussians algorithm\cite{liang2023random} is shown to be promising in comparison with the RBE for long-range interactions, and the implementation of this algorithm in the RBMD will provide a faster solution than the current version. There are other modules to be added to the RBMD, such as the TIP4P water model\cite{horn2004development}, isothermal-isobaric ensemble, and more useful force fields \cite{zeng2024constant}. Simultaneously, enabling parallel acceleration with multiple CPUs and GPUs for the RBMD is also an essential requirement in our future plan for ultra-large-scale simulations.

\section*{Acknowledgments}
This work was supported by the National Natural Science Foundation of China (Grants No. 12325113, 12071288 and 12350710181) and the Science and Technology Commission of Shanghai Municipality (Grants No. 21JC1403700 and 23JC1402300).  The authors would like to thank the support from the Center for High Performance Computing at Shanghai Jiao Tong University and the Fundamental
Research Funds for the Central Universities.

\bibliographystyle{elsarticle-num}
\bibliography{rbe_gpu_reference}

\begin{thebibliography}{10}
\expandafter\ifx\csname url\endcsname\relax
  \def\url#1{\texttt{#1}}\fi
\expandafter\ifx\csname urlprefix\endcsname\relax\def\urlprefix{URL }\fi
\expandafter\ifx\csname href\endcsname\relax
  \def\href#1#2{#2} \def\path#1{#1}\fi

\bibitem{karplus1990molecular}
M.~Karplus, G.~A. Petsko, Molecular dynamics simulations in biology, Nature
  347~(6294) (1990) 631--639.

\bibitem{Axel1987Science}
A.~T. Br\"unger, J.~Kuriyan, M.~Karplus, {Crystallographic R factor refinement
  by molecular dynamics}, Science 235~(4787) (1987) 458--460.

\bibitem{morgane2018polymer}
M.~Morgane, Z.~Zhai, M.~Perez, O.~Lame, C.~Fusco, L.~Chazeau, A.~Makke,
  G.~Marque, J.~Morthomas, Polymer chain generation for coarse-grained models
  using radical-like polymerization, Commun. Comput. Phys. 24~(3) (2018)
  885--898.

\bibitem{CiCP-33-57}
M.~Lauricella, L.~Chiodo, F.~Bonaccorso, M.~Durve, A.~Montessori,
  A.~Tiribocchi, A.~Loppini, S.~Filippi, S.~Succi, Multiscale hybrid modeling
  of proteins in solvent: {SARS-CoV2} spike protein as test case for {Lattice
  Boltzmann} — all atom molecular dynamics coupling, Commun. Comput. Phys.
  33~(1) (2023) 57--76.

\bibitem{leiserson2020there}
C.~E. Leiserson, N.~C. Thompson, J.~S. Emer, B.~C. Kuszmaul, B.~W. Lampson,
  D.~Sanchez, T.~B. Schardl, There’s plenty of room at the {Top}: {What will
  drive computer performance after {Moore’s} law?}, Science 368~(6495) (2020)
  eaam9744.

\bibitem{amber18}
T.-S. Lee, D.~S. Cerutti, D.~Mermelstein, C.~Lin, S.~LeGrand, T.~J. Giese,
  A.~Roitberg, D.~A. Case, R.~C. Walker, D.~M. York, {GPU}-accelerated
  molecular dynamics and free energy methods in {Amber18}: Performance
  enhancements and new features, J. Chem. Inf. Model. 58~(10) (2018)
  2043–2050.

\bibitem{gromacsJCP2020}
S.~Páll, A.~Zhmurov, P.~Bauer, M.~Abraham, M.~Lundborg, A.~Gray, B.~Hess,
  E.~Lindahl, {Heterogeneous parallelization and acceleration of molecular
  dynamics simulations in GROMACS}, J. Chem. Phys. 153~(13) (2020) 134110.

\bibitem{THOMPSON2022108171}
A.~P. Thompson, H.~M. Aktulga, R.~Berger, D.~S. Bolintineanu, W.~M. Brown,
  P.~S. Crozier, P.~J. In't~Veld, A.~Kohlmeyer, S.~G. Moore, T.~D. Nguyen,
  et~al., {LAMMPS} - a flexible simulation tool for particle-based materials
  modeling at the atomic, meso, and continuum scales, Comput. Phys. Commun. 271
  (2022) 108171.

\bibitem{phillips2020scalable}
J.~C. Phillips, D.~J. Hardy, J.~D.~C. Maia, J.~E. Stone, J.~V. Ribeiro, R.~C.
  Bernardi, R.~Buch, G.~Fiorin, J.~H{\'e}nin, W.~Jiang, et~al., Scalable
  molecular dynamics on {CPU} and {GPU} architectures with {NAMD}, J. Chem.
  Phys. 153~(4) (2020) 044130.

\bibitem{OpenMM2017}
P.~Eastman, J.~Swails, J.~D. Chodera, R.~T. McGibbon, Y.~Zhao, K.~A. Beauchamp,
  L.-P. Wang, A.~C. Simmonett, M.~P. Harrigan, C.~D. Stern, et~al., {OpenMM 7}:
  Rapid development of high performance algorithms for molecular dynamics, PLoS
  Comput. Biol. 13~(7) (2017) e1005659.

\bibitem{jones2022accelerators}
D.~Jones, J.~E. Allen, Y.~Yang, W.~F. Drew~Bennett, M.~Gokhale, N.~Moshiri,
  T.~S. Rosing, Accelerators for classical molecular dynamics simulations of
  biomolecules, J. Chem. Theory Comput. 18~(7) (2022) 4047--4069.

\bibitem{huang2022sponge}
Y.~Huang, Y.~Xia, L.~Yang, J.~Wei, Y.~I. Yang, Y.~Gao, {SPONGE}: {A
  GPU}-accelerated molecular dynamics package with enhanced sampling and
  {AI}-driven algorithms, Chin. J. Chem. 40~(1) (2022) 160--168.

\bibitem{wang2018deepmd}
H.~Wang, L.~Zhang, J.~Han, E.~Weinan, {DeePMD}-kit: A deep learning package for
  many-body potential energy representation and molecular dynamics, Comput.
  Phys. Commun. 228 (2018) 178--184.

\bibitem{fan2022gpumd}
Z.~Fan, Y.~Wang, P.~Ying, K.~Song, J.~Wang, Y.~Wang, Z.~Zeng, K.~Xu,
  E.~Lindgren, J.~M. Rahm, et~al., {GPUMD}: A package for constructing accurate
  machine-learned potentials and performing highly efficient atomistic
  simulations, J. Chem. Phys. 157~(11) (2022) 114801.

\bibitem{RevModPhys.82.1887}
R.~H. French, V.~A. Parsegian, R.~Podgornik, R.~F. Rajter, A.~Jagota, J.~Luo,
  D.~Asthagiri, M.~K. Chaudhury, Y.-m. Chiang, S.~Granick, et~al., Long range
  interactions in nanoscale science, Rev. Mod. Phys. 82 (2010) 1887--1944.

\bibitem{Hockney1988Computer}
R.~W. Hockney, J.~W. Eastwood, {Computer Simulation Using Particles}, CRC
  Press, 1988.

\bibitem{darden1993particle}
T.~Darden, D.~York, L.~Pedersen, Particle mesh {Ewald}: An {$N \cdot\log (N)$}
  method for {Ewald} sums in large systems, J. Chem. Phys. 98~(12) (1993)
  10089--10092.

\bibitem{essmann1995smooth}
U.~Essmann, L.~Perera, M.~L. Berkowitz, T.~Darden, H.~Lee, L.~G. Pedersen, {A
  smooth particle mesh {Ewald} method}, J. Chem. Phys. 103~(19) (1995)
  8577--8593.

\bibitem{deserno1998mesh}
M.~Deserno, C.~Holm, How to mesh up {Ewald} sums. {II. An} accurate error
  estimate for the particle--particle--particle-mesh algorithm, J. Chem. Phys.
  109~(18) (1998) 7694--7701.

\bibitem{dos2016simulations}
A.~P. dos Santos, M.~Girotto, Y.~Levin, Simulations of {Coulomb} systems with
  slab geometry using an efficient {3D Ewald} summation method, J. Chem. Phys.
  144~(14) (2016) 144103.

\bibitem{ayala2021scalability}
A.~Ayala, S.~Tomov, M.~Stoyanov, J.~Dongarra, Scalability issues in {FFT}
  computation, in: Parallel Computing Technologies: 16th International
  Conference, PaCT 2021, Kaliningrad, Russia, September 13--18, 2021,
  Proceedings 16, Springer, Association for Computing Machinery, 2021, pp.
  279--287.

\bibitem{greengard1987fast}
L.~Greengard, V.~Rokhlin, A fast algorithm for particle simulations, J. Comput.
  Phys. 73~(2) (1987) 325--348.

\bibitem{barnes1986hierarchical}
J.~Barnes, P.~Hut, A hierarchical {$O(N\log N)$} force-calculation algorithm,
  Nature 324~(6096) (1986) 446--449.

\bibitem{maggs2002local}
A.~Maggs, V.~Rossetto, Local simulation algorithms for {C}oulomb interactions,
  Phys. Rev. Lett. 88~(19) (2002) 196402.

\bibitem{pei2023fast}
R.~Pei, T.~Askham, L.~Greengard, S.~Jiang, A fast method for imposing periodic
  boundary conditions on arbitrarily-shaped lattices in two dimensions, J.
  Comput. Phys. 474 (2023) 111792.

\bibitem{arnold2013comparison}
A.~Arnold, F.~Fahrenberger, C.~Holm, O.~Lenz, M.~Bolten, H.~Dachsel, R.~Halver,
  I.~Kabadshow, F.~G{\"a}hler, F.~Heber, et~al., Comparison of scalable fast
  methods for long-range interactions, Phys. Rev. E 88~(6) (2013) 063308.

\bibitem{backus1978can}
J.~Backus, Can programming be liberated from the von {Neumann} style? {A}
  functional style and its algebra of programs, Commun. ACM 21~(8) (1978)
  613--641.

\bibitem{jin2021random}
S.~Jin, L.~Li, Z.~Xu, Y.~Zhao, A random batch {Ewald} method for particle
  systems with {Coulomb} interactions, SIAM J. Sci. Comput. 43~(4) (2021)
  B937--B960.

\bibitem{liang2021random}
J.~Liang, Z.~Xu, Y.~Zhao, Random-batch list algorithm for short-range molecular
  dynamics simulations, J. Chem. Phys. 155~(4) (2021) 044108.

\bibitem{liang2022superscalability}
J.~Liang, P.~Tan, Y.~Zhao, L.~Li, S.~Jin, L.~Hong, Z.~Xu, Superscalability of
  the random batch {Ewald} method, J. Chem. Phys. 156~(1) (2022) 014114.

\bibitem{liang2022random}
J.~Liang, P.~Tan, L.~Hong, S.~Jin, Z.~Xu, L.~Li, A random batch {Ewald} method
  for charged particles in the isothermal--isobaric ensemble, J. Chem. Phys.
  157~(14) (2022) 144102.

\bibitem{robbins1951textordfemininea}
H.~Robbins, S.~Monro, A stochastic approximation method, Ann. Math. Stat.
  22~(3) (1951) 400--407.

\bibitem{jin2020random}
S.~Jin, L.~Li, J.-G. Liu, Random batch methods {(RBM)} for interacting particle
  systems, J. Comput. Phys. 400 (2020) 108877.

\bibitem{jin2021convergence}
S.~Jin, L.~Li, J.-G. Liu, Convergence of the random batch method for
  interacting particles with disparate species and weights, SIAM J. Numer.
  Anal. 59~(2) (2021) 746--768.

\bibitem{jin2022random}
S.~Jin, L.~Li, Y.~Sun, On the random batch method for second order interacting
  particle systems, Multiscale Model. Simul. 20~(2) (2022) 741--768.

\bibitem{li2020random}
L.~Li, Z.~Xu, Y.~Zhao, A random-batch {Monte Carlo} method for many-body
  systems with singular kernels, SIAM J. Sci. Comput. 42~(3) (2020)
  A1486--A1509.

\bibitem{carrillo2021consensus}
J.~A. Carrillo, S.~Jin, L.~Li, Y.~Zhu, A consensus-based global optimization
  method for high dimensional machine learning problems, ESAIM Contr. Optim.
  Ca. 27 (2021) S5.

\bibitem{JinJCM-39-897}
F.~Golse, S.~Jin, T.~Paul, The random batch method for {$N$}-body quantum
  dynamics, J. Comput. Math. 39~(6) (2021) 897--922.

\bibitem{CiCP-28-1907}
S.~Jin, X.~Li, Random batch algorithms for quantum {Monte Carlo} simulations,
  Commun. Comput. Phys. 28~(5) (2020) 1907--1936.

\bibitem{CiCP-31-997}
J.~A. Carrillo, S.~Jin, Y.~Tang, Random batch particle methods for the
  homogeneous {Landau} equation, Commun. Comput. Phys. 31~(4) (2022) 997--1019.

\bibitem{Ewald1921AnnPhys}
P.~P. Ewald, {Die Berechnung optischer und elektrostatischer Gitterpotentiale},
  Ann. Phys. 369~(3) (1921) 253--287.

\bibitem{PhysRev.159.98}
L.~Verlet, Computer "experiments" on classical fluids. {I. T}hermodynamical
  properties of {Lennard-Jones} molecules, Phys. Rev. 159 (1967) 98--103.

\bibitem{bolstad2023vtk}
M.~Bolstad, K.~Moreland, D.~Pugmire, D.~Rogers, L.-T. Lo, B.~Geveci, H.~Childs,
  S.~Rizzi, {VTK}-m: Visualization for the exascale era and beyond, in: ACM
  SIGGRAPH 2023 Talks, Association for Computing Machinery, 2023, pp. 1--2.

\bibitem{wang2004development}
J.~Wang, R.~M. Wolf, J.~W. Caldwell, P.~A. Kollman, D.~A. Case, Development and
  testing of a general amber force field, J. Comput. Chem. 25~(9) (2004)
  1157--1174.

\bibitem{brooks1983charmm}
B.~R. Brooks, R.~E. Bruccoleri, B.~D. Olafson, D.~J. States, S.~a. Swaminathan,
  M.~Karplus, {CHARMM}: a program for macromolecular energy, minimization, and
  dynamics calculations, J. Comput. Chem. 4~(2) (1983) 187--217.

\bibitem{oostenbrink2004biomolecular}
C.~Oostenbrink, A.~Villa, A.~E. Mark, W.~F. Van~Gunsteren, A biomolecular force
  field based on the free enthalpy of hydration and solvation: the {GROMOS}
  force-field parameter sets 53{A}5 and 53{A}6, J. Comput. Chem. 25~(13) (2004)
  1656--1676.

\bibitem{jorgensen1996development}
W.~L. Jorgensen, D.~S. Maxwell, J.~Tirado-Rives, Development and testing of the
  {OPLS} all-atom force field on conformational energetics and properties of
  organic liquids, J. Am. Chem. Soc. 118~(45) (1996) 11225--11236.

\bibitem{sun1998compass}
H.~Sun, P.~Ren, J.~Fried, The {COMPASS} force field: parameterization and
  validation for phosphazenes, Comput. Theor. Polym. Sci. 8~(1-2) (1998)
  229--246.

\bibitem{senftle2016reaxff}
T.~P. Senftle, S.~Hong, M.~M. Islam, S.~B. Kylasa, Y.~Zheng, Y.~K. Shin,
  C.~Junkermeier, R.~Engel-Herbert, M.~J. Janik, H.~M. Aktulga, et~al., The
  {ReaxFF} reactive force-field: development, applications and future
  directions, npj Comput. Mater. 2~(1) (2016) 1--14.

\bibitem{behler2007generalized}
J.~Behler, M.~Parrinello, Generalized neural-network representation of
  high-dimensional potential-energy surfaces, Phys. Rev. Lett. 98~(14) (2007)
  146401.

\bibitem{schutt2018schnet}
K.~T. Sch{\"u}tt, H.~E. Sauceda, P.-J. Kindermans, A.~Tkatchenko, K.-R.
  M{\"u}ller, Schnet--a deep learning architecture for molecules and materials,
  J. Chem. Phys. 148~(24) (2018) 241722.

\bibitem{bartok2010gaussian}
A.~P. Bart{\'o}k, M.~C. Payne, R.~Kondor, G.~Cs{\'a}nyi, Gaussian approximation
  potentials: The accuracy of quantum mechanics, without the electrons, Phys.
  Rev. Lett. 104~(13) (2010) 136403.

\bibitem{thompson2015spectral}
A.~P. Thompson, L.~P. Swiler, C.~R. Trott, S.~M. Foiles, G.~J. Tucker, Spectral
  neighbor analysis method for automated generation of quantum-accurate
  interatomic potentials, J. Comput. Phys. 285 (2015) 316--330.

\bibitem{gilmer2017neural}
J.~Gilmer, S.~S. Schoenholz, P.~F. Riley, O.~Vinyals, G.~E. Dahl, Neural
  message passing for quantum chemistry, in: International conference on
  machine learning, PMLR, JMLR.org, 2017, pp. 1263--1272.

\bibitem{frenkel2023understanding}
D.~Frenkel, B.~Smit, {Understanding Molecular Simulation: From Algorithms to
  Applications}, Elsevier, 2023.

\bibitem{andersen1983rattle}
H.~C. Andersen, Rattle: A “velocity” version of the shake algorithm for
  molecular dynamics calculations, J. Chem. Phys. 52~(1) (1983) 24--34.

\bibitem{cornell1995second}
W.~D. Cornell, P.~Cieplak, C.~I. Bayly, I.~R. Gould, K.~M. Merz, D.~M.
  Ferguson, D.~C. Spellmeyer, T.~Fox, J.~W. Caldwell, P.~A. Kollman, A second
  generation force field for the simulation of proteins, nucleic acids, and
  organic molecules, J. Am. Chem. Soc. 117~(19) (1995) 5179--5197.

\bibitem{mackerell1998all}
A.~D. MacKerell~Jr, D.~Bashford, M.~Bellott, R.~L. Dunbrack~Jr, J.~D. Evanseck,
  M.~J. Field, S.~Fischer, J.~Gao, H.~Guo, S.~Ha, et~al., All-atom empirical
  potential for molecular modeling and dynamics studies of proteins, J. Phys.
  Chem. B 102~(18) (1998) 3586--3616.

\bibitem{humphrey1996vmd}
W.~Humphrey, A.~Dalke, K.~Schulten, {VMD}: visual molecular dynamics, J. Mol.
  Graphics 14~(1) (1996) 33--38.

\bibitem{stukowski2009visualization}
A.~Stukowski, Visualization and analysis of atomistic simulation data with
  {OVITO}--the open visualization tool, Modell. Simul. Mater. Sci. Eng. 18~(1)
  (2009) 015012.

\bibitem{berendsen1984molecular}
H.~J. Berendsen, J.~v. Postma, W.~F. Van~Gunsteren, A.~DiNola, J.~R. Haak,
  Molecular dynamics with coupling to an external bath, J. Chem. Phys. 81~(8)
  (1984) 3684--3690.

\bibitem{schneider1978molecular}
T.~Schneider, E.~Stoll, Molecular-dynamics study of a three-dimensional
  one-component model for distortive phase transitions, Phys. Rev. B 17~(3)
  (1978) 1302.

\bibitem{evans1985nose}
D.~J. Evans, B.~L. Holian, The {Nose-Hoover} thermostat, J. Chem. Phys. 83~(8)
  (1985) 4069--4074.

\bibitem{liang2024energy}
J.~Liang, Z.~Xu, Y.~Zhao, Energy stable scheme for random batch molecular
  dynamics, J. Chem. Phys. 160~(3) (2024) 034101.

\bibitem{rodgers2008local}
J.~M. Rodgers, J.~D. Weeks, Local molecular field theory for the treatment of
  electrostatics, J. Phys.: Condens.Matter 20~(49) (2008) 494206.

\bibitem{walker2011electrostatics}
D.~A. Walker, B.~Kowalczyk, M.~O. de~La~Cruz, B.~A. Grzybowski, Electrostatics
  at the nanoscale, Nanoscale 3~(4) (2011) 1316--1344.

\bibitem{hu2014infinite}
Z.~Hu, Infinite boundary terms of {Ewald} sums and pairwise interactions for
  electrostatics in bulk and at interfaces, J. Chem. Theory Comput. 10~(12)
  (2014) 5254--5264.

\bibitem{hu2022symmetry}
Z.~Hu, {The symmetry-preserving mean field condition for electrostatic
  correlations in bulk}, J. Chem. Phys. 156~(3) (2022) 034111.

\bibitem{gao2023screening}
W.~Gao, Z.~Hu, Z.~Xu, A screening condition imposed stochastic approximation
  for long-range electrostatic correlations, J. Chem. Theory Comput. 19~(15)
  (2023) 4822--4827.

\bibitem{predescu2020u}
C.~Predescu, A.~K. Lerer, R.~A. Lippert, B.~Towles, J.~Grossman, R.~M. Dirks,
  D.~E. Shaw, The u-series: A separable decomposition for electrostatics
  computation with improved accuracy, J. Chem. Phys. 152~(8) (2020) 084113.

\bibitem{shaw2021anton}
D.~E. Shaw, P.~J. Adams, A.~Azaria, J.~A. Bank, B.~Batson, A.~Bell,
  M.~Bergdorf, J.~Bhatt, J.~A. Butts, T.~Correia, et~al., Anton 3: twenty
  microseconds of molecular dynamics simulation before lunch, in: Proceedings
  of the International Conference for High Performance Computing, Networking,
  Storage and Analysis, Association for Computing Machinery, 2021, pp. 1--11.

\bibitem{liang2023random}
J.~Liang, Z.~Xu, Q.~Zhou, Random batch {sum-of-Gaussians} method for molecular
  dynamics simulations of particle systems, SIAM J. Sci. Comput. 45~(5) (2023)
  B591--B617.

\bibitem{plimpton1995fast}
S.~Plimpton, Fast parallel algorithms for short-range molecular dynamics, J.
  Chem. Phys. 117~(1) (1995) 1--19.

\bibitem{yao2004improved}
Z.~Yao, J.-S. Wang, G.-R. Liu, M.~Cheng, Improved neighbor list algorithm in
  molecular simulations using cell decomposition and data sorting method,
  Comput. Phys. Commun. 161~(1-2) (2004) 27--35.

\bibitem{pall2013flexible}
S.~P{\'a}ll, B.~Hess, A flexible algorithm for calculating pair interactions on
  {SIMD} architectures, Comput. Phys. Commun. 184~(12) (2013) 2641--2650.

\bibitem{kim2023neighbor}
H.~Kim, B.~F{\'a}bi{\'a}n, G.~Hummer, Neighbor list artifacts in molecular
  dynamics simulations, J. Chem. Theory Comput. 19~(23) (2023) 8919--8929.

\bibitem{irbe2022jpca}
J.~Liang, Z.~Xu, Y.~Zhao, Improved random batch {Ewald} method in molecular
  dynamics simulations, J. Phys. Chem. A 126~(22) (2022) 3583--3593.

\bibitem{moreland2016vtk}
K.~Moreland, C.~Sewell, W.~Usher, L.-t. Lo, J.~Meredith, D.~Pugmire, J.~Kress,
  H.~Schroots, K.-L. Ma, H.~Childs, et~al., Vtk-m: Accelerating the
  visualization toolkit for massively threaded architectures, IEEE computer
  graphics and applications 36~(3) (2016) 48--58.

\bibitem{wolff1999tabulated}
D.~Wolff, W.~Rudd, Tabulated potentials in molecular dynamics simulations,
  Comput. Phys. Commun. 120~(1) (1999) 20--32.

\bibitem{liang2022hsma}
J.~Liang, J.~Yuan, Z.~Xu, {HSMA}: An {O (N)} electrostatics package implemented
  in {LAMMPS}, Comput. Phys. Commun. 276 (2022) 108332.

\bibitem{mark2001structure}
P.~Mark, L.~Nilsson, Structure and dynamics of the {TIP3P}, {SPC}, and {SPC/E}
  water models at 298 {K}, J. Phys. Chem. A 105~(43) (2001) 9954--9960.

\bibitem{ryckaert1977numerical}
J.-P. Ryckaert, G.~Ciccotti, H.~J. Berendsen, Numerical integration of the
  cartesian equations of motion of a system with constraints: molecular
  dynamics of n-alkanes, J. Comput. Phys. 23~(3) (1977) 327--341.

\bibitem{Allen2017ComputerLiquids}
M.~P. Allen, D.~J. Tildesley, {Computer Simulation of Liquids}, Oxford
  University Press, 2017.

\bibitem{george2022review}
A.~George, S.~Mondal, M.~Purnaprajna, P.~Athri, Review of electrostatic force
  calculation methods and their acceleration in molecular dynamics packages
  using graphics processors, ACS Omega 7~(37) (2022) 32877--32896.

\bibitem{zou2021breaking}
X.~Zou, S.~Xu, X.~Chen, L.~Yan, Y.~Han, Breaking the von {Neumann} bottleneck:
  architecture-level processing-in-memory technology, Sci. China Inf. Sci.
  64~(6) (2021) 160404.

\bibitem{horn2004development}
H.~W. Horn, W.~C. Swope, J.~W. Pitera, J.~D. Madura, T.~J. Dick, G.~L. Hura,
  T.~Head-Gordon, Development of an improved four-site water model for
  biomolecular simulations: {TIP4P-Ew}, J. Chem. Phys. 120~(20) (2004)
  9665--9678.

\bibitem{zeng2024constant}
L.~Zeng, X.~Tan, X.~Ji, S.~Li, J.~Zhang, J.~Peng, S.~Bi, G.~Feng, Constant
  charge method or constant potential method: Which is better for molecular
  modeling of electrical double layers?, J. Energy Chem. 94 (2024) 54--60.

\end{thebibliography}







\end{document}